\documentclass[a4paper,11pt,twoside]{book}
\usepackage{ifpdf}
\ifpdf 
    \usepackage[pdftex]{graphicx}   
    \usepackage[pdftex,     
            plainpages=false,   
            breaklinks=true,    
            hidelinks,
            pdftitle="FYP Point estimators for incomplete measurements of qutrit",
            pdfauthor="Chai Jing Hao"
           ]{hyperref} 
\else 
    \usepackage{graphicx}       
    \usepackage{hyperref}       
\fi 
\usepackage{setspace}
\usepackage{quotchap}
\usepackage{amsmath}
\usepackage{amssymb}
\usepackage{booktabs}
\usepackage[numbers]{natbib}

\newcommand{\Tr}[1]{\mathrm{Tr}\left( #1 \right)}

\newcommand{\ket}[1]{| #1 \rangle}
\newcommand{\bracket}[2]{\langle #1\, | \,#2 \rangle}
\newcommand{\ketbra}[2]{|\,#1\rangle\langle #2\,|}
\newcommand{\sandwich}[3]{\langle #1\,|\,#2\,|\,#3 \rangle}
\newcommand{\I}{\mathrm{i}}
\newcommand{\D}{\mathrm{d}}
\newcommand{\Exp}[1]{\mathrm{e}^{#1}}
\newcommand{\WITH}{\quad\mathrm{with}\quad}

\title{Studies on Point Estimator for Incomplete Measurement of Qutrits}
\author{Chai Jing Hao}
\date{\today}

\begin{document}
\setstretch{1.3}
\widowpenalty=300
\clubpenalty=300
\frontmatter
\begin{titlepage}
    \begin{center}
      \setlength{\parskip}{0pt}
      {\large\textbf{NATIONAL UNIVERSITY OF SINGAPORE}\par}
      \vfill
      {\huge \bf Studies on Point Estimators for Incomplete Tomography of Qutrits \par}
      \vfill
      {\large by \par}
      \smallskip
      {\LARGE Jing Hao CHAI \par}
      \medskip
      {\large Under the supervision of \par}
      \smallskip
      {\LARGE Berthold-Georg ENGLERT \par}
      {\LARGE Hui Khoon NG \par}
      \vfill
      {\large An Honors thesis submitted in partial fulfillment for the \par}
      {\large degree of Bachelor of Science \par}
      \bigskip
      \bigskip
      {in the \par}
      {\large Faculty of Science \par}
      {\large Department of Physics, Centre of Quantum Technologies \par}
      \bigskip
      \bigskip
      \bigskip
      {\Large November 2014 \par}
      \bigskip
    \end{center}
  \end{titlepage}

\chapter{Abstract}
What can we say about the quantum system if we have incomplete information? How do we write down a statistical operator as an estimator for the physical system at hand, consistent with the incomplete information that we already possess? This thesis looks at a basic case: the case when we possess the probability values for some but not all measurement bases required to fully characterize the quantum system. Specifically, we shall consider a three-dimensional quantum system.

We will consider maximum entropy methods and ways to compute the mean quantum state, in order to calculate a state, an estimator, that is as `representative' of prior data as possible. To this end we will provide numerical methods to compute some of these estimators, and compare these methods amongst each other. We will find that these estimators perform quite closely to each other, settling the question of which method is the best.

Within the problem of incomplete measurement, there is the puzzle about the different possible future measurements to complete the measurement of the quantum system. Each of these possibilities constitutes a different way of parameterizing the quantum system. We approach this puzzle by initially not assuming that any particular future measurement might be better than others, and eventually come to the conclusion that there exists a particular set of measurement that is `best' in order to complete the measurement of the quantum system. This is a line of approach independent of existing theory in quantum measurements, yet complementary as well. It provides us with a deeper understanding of the connection of the optimal set of measurements and the quantum system.

\chapter{Acknowledgements}
I would like to express my gratitude to my supervisor Professor Berthold-Georg Englert for giving me the opportunity to learn so much over the course of the project, as well as the freedom and ideas to explore further.

I would also like to thank Assistant Professor Ng Hui Khoon for the numerous questions that sharpen my arguments and concepts throughout the project, and her various advice in writing and presenting.

Lastly, I would like to thank Teo Yong Siah for additional discussions which has helped shaped my ideas on numerical methods for the project.

It has been a fruitful and exciting journey throughout, at times slow from lack of progress, at times fast whenever a new idea appears or a numerical method yields successful results. Some turned out to be dead ends, while others have spawned yet more questions. It is a sentiment which I am sure every researcher knows very well. Yet all contribute to my understanding of the subject, of things that work, things that do not appear to work with current understanding, and other things that do not work. Such are knowledge that cannot be gained without first trying them out.

\tableofcontents
\listoffigures
\mainmatter
\begin{savequote}[200pt]
The next great era of awakening of human intellect may well produce of a method of understanding the \emph{qualitative} content of equations. Today we cannot. Today we cannot see that the water flow equations contain such things as the barber pole structure of turbulence that one sees between rotating cylinders. Today we cannot see whether Schr{\"o}dinger's equation contains frogs, musical composers, or morality --- or whether it does not.
\qauthor{---R. P. Feynman}
\end{savequote}
\chapter{Quantum states, Tomography, and Statistical inference}
\chaptermark{Describing quantum states}
The main purpose of this thesis is to study different possible methods of estimating the state of a quantum system when we not performed sufficient measurements to fully reconstruct the state of the system. The study of this methods involve their rationale, as well as theoretical and numerical methods to calculate and compute them. We will compare these methods against each other and attempt to say which is `better'. Through the course of such a study, we also gain some insights about the structure of the space of quantum states, as well as some insights about the measurements that is used for state reconstruction. Specifically, we will look at the case when the quantum system is three dimensional (also known as a qutrit), so as to be able to apply these analyses to a slightly less simple scenario.

In this first chapter, we will set up some basic ides about quantum states, the purpose and concept behind tomography, as well as illustrating some structure of the state space that goes with the problem of incomplete tomography that is the subject of this thesis. 

\section{Describing quantum systems}
Our knowledge of physical systems is characterized in various ways depending on what we are able to know, and what is relevant to us. This knowledge is also termed the \emph{state} of the system. We might be concerned with the dimensions of a physical object, or the position and velocity of a particle. Or we might be interested in the probability of the coin coming up heads in the next toss. In quantum systems, what we can know are the probabilities for measurement outcomes. We have full knowledge of the quantum system if we are able to predict the probabilities for any kind of measurements we are to make on system.

What we can measure about the system (observables) are written as written as pure states $\ket{\hphantom{\psi}}$, with $d$ outcomes associated with states $\ket{\psi_1},\cdots,\ket{\psi_d}$. The total number of mutually exclusive outcomes that we can observe is the dimension of the system. However, what differentiates between quantum systems and classical systems are that linear superpositions of pure states are possible states as well. It is therefore possible that if we have some state $\ket{\phi}$, we can measure in terms of another set  of basis states such that 
\begin{equation}
	\ket{\phi} = \sum_{j=1}^d c_j\ket{\psi_j}\,,
\end{equation}
then the probabilities for the measurement outcomes $\ket{\psi_j}$ are $|c_j|^2=|\bracket{\phi}{\psi_j}|^2$. This is also known as Born rule for calculating probabilities for measurement outcomes.

While what we have described is for pure quantum systems, we can, in general, have classical type of statistics as well.  Adopting an operator notation similar to how other observables like Hamiltonian are treated, we can instead write the same pure quantum state as $\ketbra{}{}$. Now we can consider convex mixtures of more pure states like
\begin{equation}
	\rho = \sum_j p_j\ketbra{\psi_j}{\psi_j}\,.
\end{equation}
In particular, we can have a classical mixture with orthogonal states, or in classical parlance, mutually exclusive measurement outcomes. The resulting statistics of the measurement of the states in this particular basis would be classical. The probability of finding the system in pure state $\ketbra{\phi}{\phi}$, is now given by the more general expression $\Tr{\rho\ketbra{\phi}{\phi}}$. Arguably, the Born rule is the main equation on which much of the theory on quantum measurements and state estimation is based. It is simple, but how would we know how much depth there is in the equation?

In more simplistic descriptions, a direct measurement on the system involves eigenstates of an observable, which constitute a basis for the system. This type of measurement is also known as von Neumann measurement or projective measurement. Through many observations of measurement outcomes for many identitical copies of the quantum system, we build up statistics that point us to the probabilities of the outcomes with which we can record. 

\section{Tomography}
Calculating probabilities for measurement outcomes given a state is straightforward; doing the reverse is not quite so straightforward. To illustrate this difficulty, consider a qubit state $\ket{\uparrow_z}$. It has equal probability of being found in $\ket{\uparrow_x}$ or $\ket{\downarrow_x}$ in an experiment set up to measure observable $\sigma_x$. However, the state $\ket{\uparrow_y}$ gives the same probabilities for the same measurement. Given just probabilities for one von Neumann measurement does not allow us to fully deduce the state of the system. Moreover, after a measurement is performed, the original state of the system is destroyed completely. Hence, we need to perform several measurements for different bases, using several identical copies of the state in order to obtain full information about the system. Hence, the term \emph{tomography} for this process. The full tomography of the system involves us learning enough parameters to explain all possible statistics that we are able to get from measuring the physical system. And this involves the more general statistical operator, and its parameterization.

\subsection{Parameterizing statistical operator}
The statistical operator of dimension $d$ requires $d^2-1$ numbers to write down. Complete specification of the quantum system means that we are able to write down the probabilities of observable outcomes for any measurement for that system. Amongst the different ways of parameterizing the statistical operator, we might suppose that we will learn about the system via a set of bases for the Hilbert space. We can write down the state of the system as
\begin{equation}
  \label{eq-basesparameterize}
  \rho - \frac{\mathbb I}{d} = \sum_{j=1}^{d+1} \sum_{k=1}^{d} c_{jk} \Lambda_{jk} \,,
\end{equation}
where $\Lambda_{jk} = \ketbra{e^{(j}_k}{e^{(j)}_k} - \mathbb I/d$. That is, we have expressed the traceless set of hermitian matrices as a linear combination of $d+1$ bases (not necessarily mutually unbiased) each consisting of $d$ orthonormal kets. We determine the coefficients $c_{jk}$ by the Born rule,
\begin{equation}
  \label{eq-probabilityreconstruct}
\Tr{\rho \ketbra{e^{(l)}_m}{e^{(l)}_m}} - \frac{1}{d} = \sum_{jk} c_{jk}\Tr{\Lambda_{jk} \ketbra{e^{(l)}_m}{e^{(l)}_m}}\,,
\end{equation}
which stands for a set of linear $d(d+1)$ equations. Henceforth, we shall count the double indices $(j,k)$ with one index $k$, and $(l,m)$ with $j$. We shall write the matrix with elements $\Tr{\Lambda_{jk}\ketbra{e^{(l)}_m}{e^{(l)}_m}}$ as $M$.
However, there are additional constraints placed on the left hand of the equation. In particular, the probabilities measured in each basis have to sum to one. 
\begin{equation}
	0 = \sum_{j=a}^{j=a+d-1} \sum_k M_{jk} c_k\,,
\end{equation}
for $a =1,1+d,1+2d,\cdots,1+d^2$. 
As a result, some columns of $M$ are linearly dependent. We can solve for the coefficients $c_k$ by multiplying the pseudoinverse $M^+$ to the column of probabilities $p_k$. The reason for this is because we are trying to write $\rho$ as a $d(d+1)$ vector even though $\rho$ is only parameterized by $(d-1)(d+1)$ parameters. We must remember that only a subset of probabilities correspond to $\rho$ with non-negative eigenvalues. Why do we not choose to parameterize $\rho$ in only $d^2-1$ parameters? The answer is that since we can only obtain frequencies from experiments which approximate probabilities, our data has constraints. As is usually the case with real experimental data, the data may be unable to fulfill all constraints due to imperfections of detectors. In this case it is not immediately clear which number amongst the $d$ probability values for a basis to eliminate in order form a complete vector basis $\{\Lambda_i\}$, so we have to work with an overcomplete set. Such a set is not called a basis for a vector space, but rather a \emph{frame}. 

For concreteness, let us consider the case of the qutrit. That means $M$ is a $12\times 12$ matrix, and we require 12 probabilities for 4 sets of bases in order to reconstruct the state. 

\section{Incomplete tomography and statistical inference}
\sectionmark{Incomplete tomography}
We consider, as a beginning point, the situation where tomography of the system is incomplete. Such a situation is more physically relevant than the ideal situation where we are in possession of all information about the system. Due to finite resources, an experimenter may not be able to have a large enough observation sample to say with confidence the probability of a measurement outcome; he/she may have failed to measure enough bases to fully reconstruct the state; otherwise the information obtained may be in summarized form such as the expectation value of an observable like the Hamiltonian for a hydrogen atom. We want to be able to infer as much as we can in such situations of incomplete information, much like how we are still able to make predictions about a gas or the weather even when we have insufficient information to completely describe every aspect of the system.

Informationally incomplete quantum tomography has been studied in simpler theoretical contexts in \cite{wiedemann1999quantum}, and some numerical procedures reviewed in \cite{teo2012incomplete}. In this project, we will consider in more detail the inference process. We focus on the procedure for writing down the statistical operator for incomplete tomography on a system and do not consider other details such as the context of the physical situation (photon count, atomic levels, etc). One of the trickier problems in quantum state estimation in general is that the constraint of the positivity of the statistical operator restricts the observable probabilities of measurement outcomes to a convex subset. Equation \eqref{eq-basesparameterize} is not surjective; while we can find unique probabilities for measurement outcomes for every state, we cannot take any set of probabilities and expect to find a state. The boundary of the convex subset of probabilities is described by $\lambda_\text{min}=0$ for the minimum eigenvalue. It consists of states such that each of them is extremal, somewhat like the surface of ball. The geometry of the space of states is therefore difficult to study since the boundary does not admit a simpler description other than in terms of the minimum eigenvalue. Only in the qubit case can we describe the boundary in simple geometry: as the Bloch sphere. For the qutrit, the boundary is a 7-dimensional closed hypersurface, which makes it hard to study. However, by making some measurements and obtaining partial information about the qutrit, we can study the geometry of the state space in a reduced dimension space where some information is now fixed by prior measurements. In the remainder of the report, we will be able to see figures that illustrate the geometry of the state space in reduced dimensions. Through the study of incomplete tomography, we may gain some insight about the state space and its connection to measurements.

Let us start with a simpler scenario for the qutrit, and suppose now we have measured 3 out 4 bases, that means we have 9 probabilities in hand, and 3 undetermined. The first order of business is to determine the set of probabilities that are consistent with the probabilities we have already measured, and also ensure that the density matrix is non-negative. In this scenario, if the 3 measured bases come from the set of mutually unbiased bases (MUB)\footnote{See appendix for an explanation about the MUB.}, then we can express the density matrix in terms of the last basis from the MUB set that is unmeasured. In this basis, the unknown elements are on the diagonal of the 3 by 3 density matrix, whereas the off-diagonal elements are precisely known. This simple case involving 3 unknown probabilities or simply just 2 independent variables allows us to put some of our results to be illustrated on paper, which aids visualization. But when we  talk about more general theory, we may instead refer to a more general setting with more bases unmeasured; in this case, the prior information would be probabilities of measured bases, and our task is to estimate probabilities for unmeasured ones. 

Put more formally, we have $\rho$ as a function of unmeasured probabilities $\vec p$ in
\begin{equation}
	\rho(\vec p) = \frac{1}{3}\mathbb I + \sum_{i=1}^{12} c_i(\vec p;\,\vec f)\,\Lambda_i\,,
\end{equation}
where we have the constraints that 
\begin{equation}
	f_j = \Tr{\rho \Lambda_j}
\end{equation}
with $f_j$ as the observed frequencies of the outcomes for measured bases $\Lambda_j$. In general, the frequencies contain statistical fluctuations, but we shall assume for the moment, that the frequencies are exact probabilities. The coefficients $c_i$ depends on both $\vec p$ and $\vec f$. We also should not forget the constraint that 
\begin{equation}
	\rho(\vec p) \geq 0\,.
\end{equation}
If we have already measured 3 out 4 bases required to parameterize the qutrit, then we can write the density matrix in terms of the last unmeasured basis. Then the problem will be in the form
\begin{equation}
	\label{eq-begintask}
	\rho(x,y,z) = \begin{pmatrix} x & a & b\\ a^* & y & c \\ b^* & c^* & z \end{pmatrix}
\end{equation}
such that $a$, $b$, $c$ are known complex numbers, and $x$, $y$, $z$ are unknown to us. In this form, the computational basis is mutually unbiased with respect to all the other measured bases, but currently there is no \emph{a priori} reason other than symmetry based on mathematical theory for the MUB, for an experimenter to choose to measure in this basis in the future.

\begin{figure}
	\centering
	\includegraphics[width=0.8\textwidth]{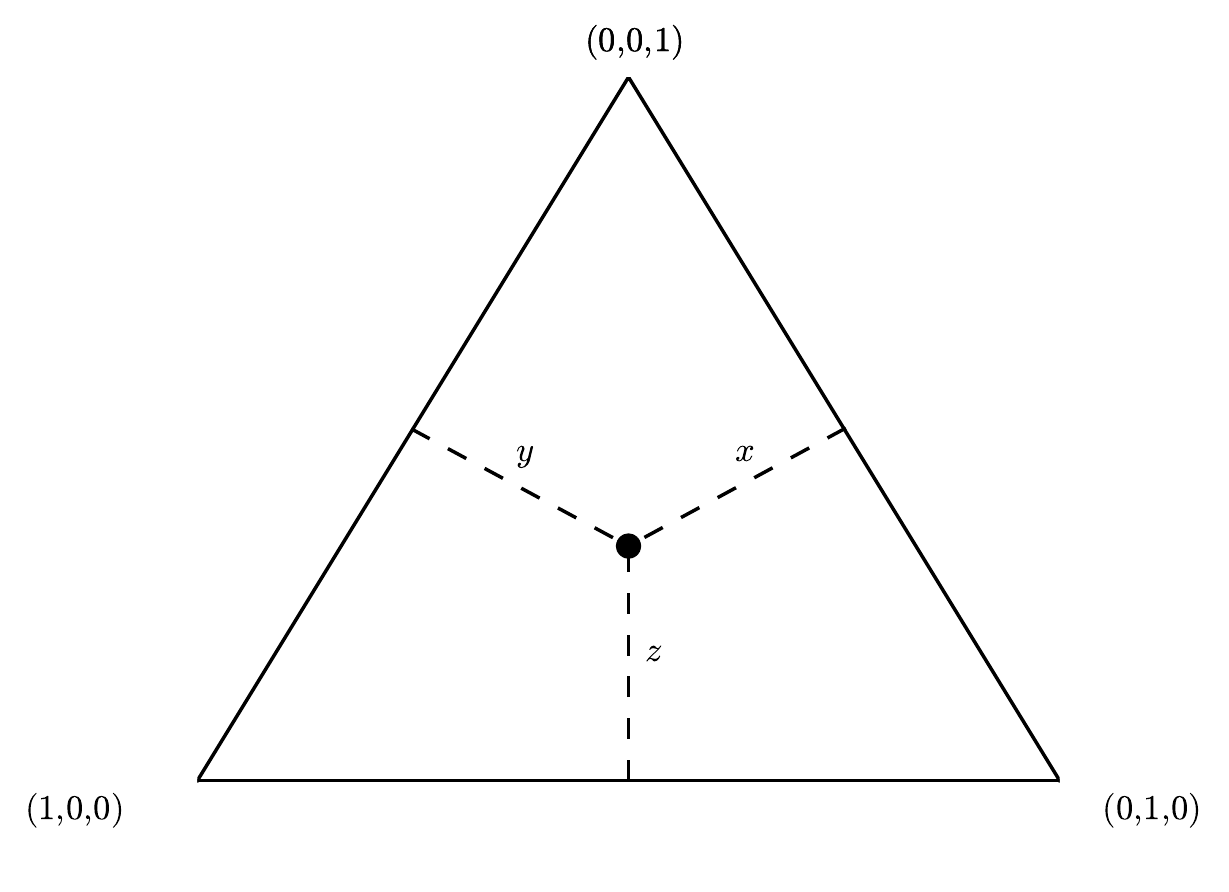}
	\caption{Representing probabilities on a simplex. For 3 probabilities, the simplex is an equilateral triangle. A point $(x,y,z)$ is represented accordingly by the perpendicular distance to the sides of the triangle, or simply by rewriting the vector in terms of 2 linearly independent vectors.}
	\label{fig-paramtriangle}
\end{figure}

When we only have one basis unmeasured, states that are compatible with the measured data can be parameterized by three probabilities $\rho(x,y,z)$ as in \eqref{eq-begintask}. We can visualize the states on the probability simplex as in figure \ref{fig-paramtriangle}.
\section{Permissible region}
\begin{figure}
  \centering
  \includegraphics[width=0.8\textwidth]{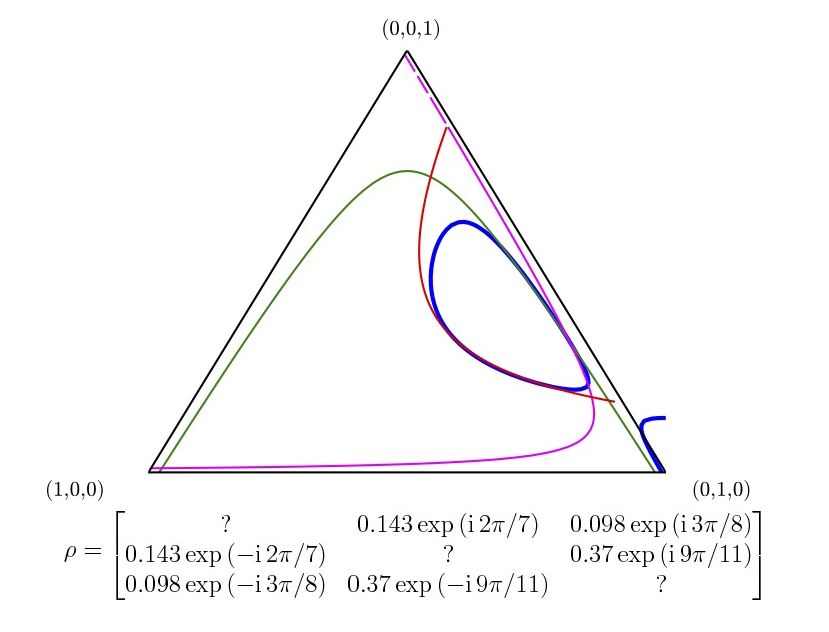}
  
  \caption{Illustration of the non-negative principle sub-determinants as condition to find the permissible region. The blue curve is the condition that determinant of $\rho$ to be non-negative. The green, red, magenta curves the condition that the 2 by 2 principle subdeterminants be nonnegative. And the probability simplex enforces the condition that the 1 by 1 principle sub-determinants be non-negative.} 
  \label{fig-boundaryequation}
\end{figure}

Under the constraint of fixed probabilities for some measurement bases, the permissible region\footnote{which we shall call $\Gamma$} of probabilities for the remaining unmeasured bases can be determined in various ways. One is by Cholesky factorization of $\rho$, which is the fastest numerical method. A more analytical approach is to consider the condition that all principle sub-determinants of $\rho$ must be non-negative. It was found that the boundary of the permissible region is described by the algebraic equation $\det \rho = 0$. Since the determinant is a general polynomial with closed and open branches, the correct arm of the polynomial is selected by the non-negativity condition of the other sub-determinants of lower order (figure \ref{fig-boundaryequation}).  

\begin{figure}
  \centering
  \includegraphics[width=0.8\textwidth]{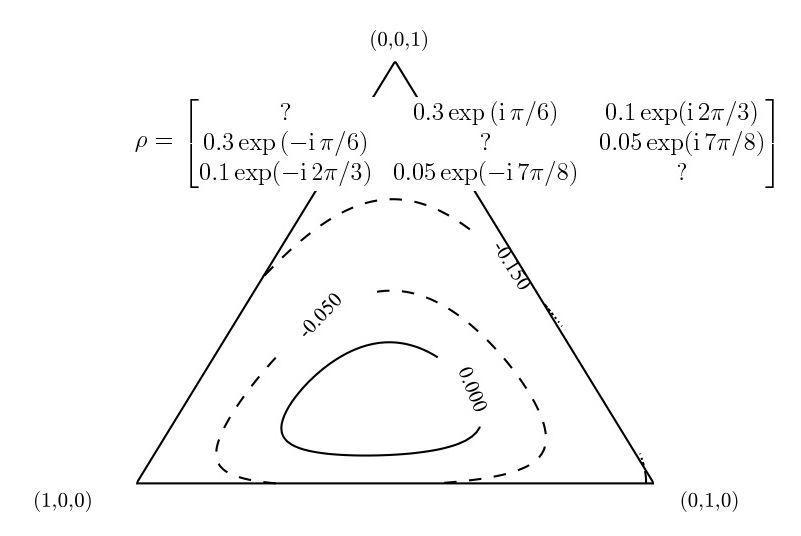}
  
  \caption{Mapping out the minimum eigenvalue $\lambda_\text{min}$ for the density matrix corresponding to the probability values. The region where $\lambda_\text{min} \geq 0$ is the permissible region.}
  \label{fig-mineig}
\end{figure}

It is my opinion however, that a more useful approach is to use minimum eigenvalue of $\rho$ (figure \ref{fig-mineig}). Through many numerical experiments, it was observed that the general sizes and locations of these permissible regions can greatly vary depending on the known values --- from regions that almost include the whole probability simplex, to small ones that are almost point-like. While the Cholesky factorization is numerically fast, calculating the minimum eigenvalue is useful in cases when the permissible region has small area relative to the simplex. In these cases, locating the region through random Monte Carlo or through a discrete mesh of the simplex is inefficient, or could even miss the permissible region altogether. In these cases, by using minimum eigenvalue one can perform optimization to reach the permissible region. With an efficient gradient ascent algorithm, this can be computed with relative ease. 

Any estimator that is consistent with prior measurements has to lie within the permissible region. As we can see, the permissible region's analytical description is not very useful in actual computation, so numerical computation is necessary, and these algorithms have to take into account this constraint.

\begin{savequote}[200pt]
But this long run is a misleading guide to current affairs. In the long run we are all dead. [Theorists] set themselves too easy, too useless a task if in tempestuous seasons they can only tell us that when the storm is long past the ocean is flat again.
\qauthor{J. M. Keynes}
\end{savequote}
\chapter{Inference}
Statistical inference is applicable independent of whether we are working with quantum systems or not. Rather, once we have made use of the Born rule to compute probabilities of measurement outcomes, we can apply statistical inference theory to observations of the random variables.

Suppose we are trying to determine the statistical properties of a 6-sided die. Suppose further a friend has thrown the die a few times and recorded the frequency each side was observed. This friend tells you the frequencies for the first four sides but for one reason or another, withholds information about the last two sides. Can we still say something about the die? In fact, we are still able to make statements about the die, through some assumptions. Such a process of inference is also known as estimation. We will discuss some of these assumptions pertaining to statistics, while keeping in mind that assumptions are just that: assumptions. Like axioms in a mathematical theory, they cannot be justified by the theory itself but only through what we accept as reasonable. In physics, we accept it if experience and experiments show that they are facts, or if their consequences are reasonable. One of the aims of this project is to compare if adopting these assumptions lead to acceptable consequence. 

As a prelude, we will start with a definition of probability. Hitherto, we have been rather careless in how we use this term. Probability is our belief about the frequency the event will occur if a very large number of identical situations leading to the event were to take place. If I say that ``There is a 0.5 probability that the coin will come up heads in a toss", I mean that I expect that in a very very large number $N$ of identical situations, I will observe heads for $0.5N$ times. Another person may believe differently, for his assessment may be different. Probabilities are \emph{my} expecations and \emph{my} beliefs, and may differ from person to person. But what about the coin? Surely the coin is in some concrete statistical state? It could be that the coin has some concrete state independent of our observations, but we do not \emph{know} this. What we do know about the coin is what we can measure about it. Unfortunately in statistics, we can never quite obtain fully, complete information about the state. In order for me to believe that the coin is perfectly fair, I would need to observe an infinite number of tosses. Any finite number $N$ of tosses only convinces me of the probability for heads to within some finite interval. But if we are already tossing the coin an infinite number of times, we do not need to make any predictions for it. On the other hand, when we have no observations for the coin, we would like to have a prediction. But since there are no observations, our predictions would be severely limited. Probabilities are our beliefs, our expectations. Frequencies are actual data we observe. With large numbers of data, frequencies would be near the true statistical parameters of the system, but would never reveal it exactly. 
Typically in real experiments, we simply observe and record the number of times of the event occuring, and we form some estimate about the relevant statistical parameters. As a consequence of the subjectivity of the definition of probability, we find of course that the statistical operator becomes subjective. This is perfectly in line with our ideas about quantum state estimation. Rather than trying to decide whether the system possesses some concrete physical `reality', we simply record and measure what we can observe and make suitable estimations where we need them. Therefore our view is that the statistical operator or wave function (in the language of Schr\"{o}dinger's wave equation) does not hold any more physical meaning other representing our best knowledge of the system at a particular instant in time.

As a side note, we note that we should really talk about finite regions of uncertainty of the probability values since we never have infinitely many observations of the system. However, it is usually convenient to pick out a point in this region as the representative. Such a point is termed the point estimator, and hence people use statements such as ``The probability for the coin to toss heads is 0.5" and not statements such as ``The probability for the coin to toss heads is $[0.48,0.52]$".

In the following sections, we will briefly look at some of the (sometimes competing, sometimes complementary) assumptions in statistical inference theory. A more detailed overview can be found in \cite{jaynes1979we}.

\section{Principle of indifference}
In order to solve a problem of prediction such as

\begin{center}
\parbox[c][4em][c]{0.6\textwidth}{\centering What is the probability that there will be 31 heads in 50 tosses of the coin?}
\end{center}

\noindent
requires at least an initial assignment of probability values. In this case we need to decide first on

\begin{center}
\parbox[c][4em][c]{0.6\textwidth}{\centering What is the probability that the the coin will land heads in the next toss?}
\end{center}

\noindent
If this value is not specified, then we need to estimate it. One assumption that is adopted is to say: Since we have no \emph{a priori} reason to believe any particular outcome is favored, unless there is evidence otherwise, we say that the probability of each outcome of the coin is equal. This is the principle of indifference, sometimes also known as principle of equiprobability, first stated as an explicit formal principle in Ars Conjectandi by Jacob Bernoulli (1713). It is straightforward to apply when there is finite and discrete number of outcomes of a random variable, but is less so when the outcomes are continuous.

\section{Bayes Rule}
Having observed a coin land heads 31 times out of 50 could help us decide the probability of the coin landing heads in a next toss. The probability that the probability $p$ for heads lies within an interval $p$ and $p+\D p$ was worked out by Thomas Bayes (1763), a British clergyman. It is 
\begin{equation}
	\frac{51!}{31!19!}p^{31}(1-p)^{19}\,\D p\,.
\end{equation}
We recognize in modern terms, $\frac{(n+m+1)!}{n!m!}p^n(1-p)^m$ as the likelihood function for the coin. Later on, Laplace in his 1774 memoir (probabilities of causes), stated that if we have a conceptual model for an event $E$ and its set of conceivable causes $C_i$, and if we have also have obtained some observations of the event and now wish to infer which $C_i$ as the cause for the event, then 
\begin{equation}
	\mathrm{prob}(C_i|E) = \frac{\mathrm{prob}(E|C_i)\mathrm{prob}(C_i)}{\sum_j \mathrm{prob}(E|C_j)\mathrm{prob}(C_j)}
\end{equation}
which in modern literature is now called Bayes rule. In more modern parlance, $\mathrm{prob}(E|C_i)$ would the likelihood, while $\mathrm{prob}(C_i|E)$ is the posterior. The proportionality factor converting one to the other is the prior, $\mathrm{prob}(C_i)$, the probability for causes. The prior cannot be decided by experiment, but rather is the very initial assignment of probability values that is required before we can start to calculate anything. While Laplace originally sought to formulate the ``probability of causes" as a way to avoid the principle of indifference because we are not always able to break things down into an enumeration of equally possible cases, he found that it came back again in the form of the prior. In particular, while Bayes rule tell us how to update our state of knowledge when receiving new information, it does not tell us how to start our initial state of knowledge. That is arbitrary. While it is true that in the long run of many observations, these pieces of information render the initial state of knowledge useless, we are usually confronted with the middle situation of having obtained some knowledge, but it is incomplete.

\section{Jeffreys prior}
If a probability distribution needs to be assigned to a continuous space, then it is not clear exactly how to apply the principle of indifference. It is insufficient to specify a uniform density, because this is ambiguous due to a lack of invariance under a change of parameters. A uniform density for a parameter $x$ will not be uniform for $y=x^3$ or $z=\exp{x}$. In 1945, Sir H. Jeffreys proposed in ``An invariant form for prior probability in estimation problems" that the prior could be selected under all possible changes of parameters. Suppose that a probability could be parameterized by $p(\vec x)$. By considering the infinitesimal distance between probability distributions using Kullback-Leibler divergence $\mathbb{E}_{\vec x} \log\frac{p(\vec x)}{q(\vec x)}$ and Hellinger distance $\int (\sqrt{\D p}-\sqrt{\D q})^2$, and approximating to second order, we can observe that both lead to a common expression 
\begin{equation}
	\mathbb{E}_{\vec x} \left[\frac{\partial \log p(\vec x)}{\partial x_i}\frac{\partial \log p(\vec x)}{\partial x_j} \right]
\end{equation}
which we now call the Fisher information matrix $I(x)$. Under a change of parameters from $\vec x$ to $\vec y$,
\begin{equation}
	I(\vec y)=J^\mathrm{T}I(\vec x(\vec y))J\,,
\end{equation}
where $J_{ij} = \frac{\partial x_i}{\partial y_j}$ is the Jacobian matrix.
By defining the prior as $\sqrt{\mathrm{det}\,I}$, the prior becomes invariant under reparameterization since the Jacobian factors cancel out with the derivatives from the new information matrix. This prior is also known as Jeffreys prior or the non-informative prior. However, there is some arbitrariness in deriving the Fisher information matrix from the probability distances. We can see a familiar connection between the Jacobian and the square root determinant of the metric based on concepts for a manifold (in this case, a statistical manifold), but this connection also tell us that the measure for the metric can be arbitrary. As is noted in \cite{shang2013optimal}, other kinds of \emph{form}-invariant priors can be constructed by considering different kinds of measures of distances. In short, the problem of the prior is still plagued by the problem of choosing a measure for the parameters of the statistical model.

\section{Maximum entropy}
In statistical mechanics, a similar problem of inference was solved by Maxwell and Boltzmann. There, we wish to find the probability distribution for the momentum of molecules in a conservative force field, under the constraint of constant total energy and total particle number. Because of this constraint, not all positions and momenta are equally likely. At this point, by dividing the phase space into discrete regions such that the energy of molecule occupying such a region does not vary much. The total number of such discrete regions is finite because of the energy constraint. In each region $R_k$, a large number $N_k$ of molecules can occupy it. For a given set of occupation numbers $\{N_k\}$, the total number of ways to realize this is 
\begin{equation}
	\Omega=\frac{N_!}{N_1!N_2!\cdots N_m!}
\end{equation}
with total energy $E = \sum_{k=1}^m N_k E_k$. Now we ask the question, amongst the different sets $\{N_k\}$ for which the constraints are satisfied, which is the most likely? At this stage, we may make use of the Stirling approximation to simplify to
\begin{equation}
	\log \Omega = -N\sum_{k=1}^m \frac{N_k}{N}\log\frac{N_k}{N}\,,
\end{equation}
where we recognize the form of what we now call \emph{entropy}. By treating $N_k/N$ as probability, which is true in the limit of a very very large number $N$, most certainly true for the case of a gas, we connect the idea of entropy with the number of ways a system can realize a particular configuration. In order to find the most likely configuration we can maximize $\Omega$ which is equivalent to maximizing $\log \Omega$. The solution by Lagrange multipliers is simple, resulting in 
\begin{equation}
	N_k = \frac{N}{Z(\beta)}\exp(-\beta E_k)
\end{equation}
where $Z(\beta)$ is known as the partition function, a normalizing factor chosen such that the energy constraint is satisfied. In this way we obtain the probability of a particle having some momentum and position. As an independent application, Shannon, in trying to characterize random messages sent over a communication channel, came to the same conclusion about maximizing entropy, in the classic paper ``A mathematical theory of communication", 1948.

What is entropy? As Jaynes describes clearly in \cite{jaynes1979we}, entropy is a measure of information possessed by a person who only has values for macroscopic quantities instead of the knowledge of the micro configurations of the systems. Maximizing the entropy in order to find a probability distribution for a system is thus quite similar to assigning a probability distribution when we have incomplete information. Hence the principle of maximum entropy can be seen to be a kind of generalization of the principle of indifference. However there is again the problem of measure: How do we know that the discrete regions of the phase space must be cut up in that manner? We nonetheless feel that it is physically correct and that this is the simplest way to do this without incurring inconsistencies.

\subsection{von Neumann entropy}
The extension of the concept of entropy begins by consideration of the mixture of orthogonal states. For example we might say of a qubit system that it has probability $p$ of putting out state $\ketbra{\uparrow_z}{\uparrow_z}$ and probability $1-p$ of putting out state $\ketbra{\downarrow_z}{\downarrow_z}$. As quantum system, there are in fact many different ways to write such a system as a mixture of pure states, but there is only one unique way of writing it as a mixture of \emph{orthogonal} pure states\footnote{as long as the eigenvalues are not degenerate} (can easily see this from uniqueness of eigenvalue decomposition). If the eigenstates of the system has been fixed, then we might consider the system as random variable involving the different mutually exclusive outcomes. Then naturally, we can talk about the entropy of this random variable. It is
\begin{equation}
	S(\rho) = -\sum_i \lambda_i\log\lambda_i=-\Tr{\rho\log\rho}\,,
\end{equation}
where $\lambda_i$ are the eigenvalues of $\rho$. The von Neumann entropy is maximized when the system is in a fully mixed state, that is for the situation when we equally expect each eigenstate to be produced.

When instead of a statistical operator $\rho$, we are given expectation values to a set of operators $\mathcal O$, we may do as Jaynes prescribes \cite{JaynesInfoStats} --- maximize the entropy subject to the constraints of the measured expectation values. That is we need to have
\begin{subequations}
	\label{eq-qentmax}
	\begin{align}
		\Tr{\rho} &= 1\,,\\
		\rho &\geq 0\,,\\
		\Tr{\rho\mathcal O_i} &= o_i\qquad \text{for } i=1,2,\cdots,n
	\end{align}
\end{subequations}
satisfied by the set of states $\Gamma$\footnote{In our parameterization \eqref{eq-basesparameterize}, these constraints are taken into account by the permissible region $\Gamma$.}, in which we desire
\begin{equation}
	\label{eq-qentmax2}
	\hat\rho = \operatorname*{arg\,max}_\Gamma \,S(\rho)
\end{equation}
which is theoretically the maximum von Neumann entropy estimator (MvNE). The theoretical solution proceeds in the same manner, by use of Lagrange multipliers,
\begin{align}
	\hat\rho &= \frac{1}{Z}\exp \left(-\sum_j \mu_j\mathcal O_j\right)\\
	Z &= \Tr{\exp\left(-\sum_j \mu_j\mathcal{O}_j\right)}\,.
\end{align}
However, $\mathcal O$ are usually non-commuting, so these equations, while being compact and illustrating the connection with statistical mechanics, are not too useful in terms of numerical computations. We will look at some numerical algorithms to compute these estimators in the following chapter. 

\subsection{Shannon entropy}
However, another way to consider our measure of ignorance of the state is to take an ensemble of identically prepared systems, choose a measurement, and perform it on the ensemble for each system. We might ask, which configuration of outcomes for the ensemble is most likely. This would also give us information about the system. Shannon entropy is defined as
\begin{equation}
	H(\rho,\{\Pi_j\}) = -\sum_j \Tr{\rho\Pi_j}\log\Tr{\rho\Pi_j}\,,
\end{equation}
where $\{\Pi_j\}$ is a measurement basis with orthogonal states. Shannon entropy is clearly applicable in the case of mutually exclusive measurement outcomes, while it is not clear of its applicability when the measurement outcomes are not mutually exclusive, such as when a probability operator measurement\footnote{But by viewing the POM as a projective measurement in a larger dimension, we can still use Shannon entropy. The price to pay however, is that we need to introduce extra degrees of freedom and entangle the original system with these extra degrees of freedom.} (POM) is used. It can be shown that the relationship between von Neumann entropy and Shannon entropy is
\begin{equation}
	\operatorname*{arg\,max}_\Gamma S(\rho) = \operatorname*{arg\,max}_\Gamma\,\operatorname*{min}_{\{\Pi_j\}}\,H(\rho,\{\Pi_j\})
\end{equation}
whereby we see that von Neumann entropy is equivalent to as if we are able to conduct the `best' measurement for each $\rho\in\Gamma$. Clearly, this is not practical in a real physical setup. We usually set up one measurement and hope to identify the system through observation of the frequencies of the chosen measurement outcomes.

However, we have to pick out a particular measurement if we want to maximize Shannon entropy. Therefore some questions we need to deal with are: Can we optimize our future measurement? If each future measurement is possible, should we assign a prior probability distribution to the space of measurement bases, and take the average? Failing to do all those, can we instead assume the worst case scenario and try to maximize entropy for this scenario? These are questions that we will also attempt to answer in this project.

\begin{savequote}[200pt]
Entropy is used as a criterion for resolving the ambiguity remaining within the problem, when we have stated all the conditions that we are aware of.
\qauthor{--- E. T. Jaynes}

If we have no information relevant to the actual value of the parameter, the probability must be chosen so as to express the fact that we have none.
\qauthor{--- H. Jeffreys}
\end{savequote}
\chapter{Estimators}
The primary difficulty with the numerical computation of the estimators is the various constraints that we need to take care of: positivity of density matrix, and fixed probability values for already measured bases. However, since the permissible region is convex, and we are maximizing concave functions (or minimizing a convex function), we can turn to well-established techniques in convex optimization \cite{boyd2004convex}. However, when we need to calculate the mean of some quantity that depends on the permissible region $\Gamma$, we then require Monte Carlo techniques. We will now discuss some ideas involved in the numerical computation of the various estimators.

\section{Some numerical preliminary}
\subsection{Gradient ascent}
With reference to figure \ref{fig-mineig}, any estimator we produce is going to have to lie within the convex permissible region $\Gamma$. We would like to be able to find an approximate location to this region without having to pick random points blindly from the simplex and testing whether the corresponding density matrix is positive. With a simple gradient ascent, we can make use of the local information of the minimum eigenvalue in order to guide infeasible iterates to the permissible region.

Amongst the set of states $\Gamma$ parameterized as
\begin{equation}
	\rho = \frac{1}{3}\mathbb I + \sum_{j} c_j\Lambda_j\,,
\end{equation}
we find that variation of the minimum eigenvalue is simply
\begin{equation}
	\delta\lambda_\text{min} = \Tr{\ketbra{E_\text{min}}{E_\text{min}}\delta\rho}\,,
\end{equation}
where $\ket{E_\text{min}}$ is the eigenvector associated with the minimum eigenvalue. The variation of the minimum eigenvalue with respect coefficients $c_j$ is then
\begin{equation}
	\delta\lambda_\text{min} = \sum_j \Tr{\ketbra{E_\text{min}}{E_\text{min}}\Lambda_j}\,\delta c_j\,.
\end{equation}
But since the coefficients $c_j$ are related to the probability values of the unmeasured bases by a linear equation,
\begin{equation}
	p_j^{(\text{unmeas})} - \frac{1}{3} = \sum_k \Tr{\Lambda_k\Lambda_j^{(\text{unmeas})}}c_k
\end{equation}
in which the coefficients $c_j$ can be solved by taking the pseudo inverse of matrix $\Tr{\Lambda_j\Lambda_k}$. In the linear inversion process, both measured and unmeasured probabilities go into the linear equation, as the coefficients themselves are properly a function of both measured and unmeasured probabilities. Therefore in this way we can work out the the variation of the minimum eigenvalue with respect to unmeasured probabilities. However, since optimization on the domain of probabilities is still subjected to the unit probability constraint, we remove this by rewriting the probabilities as a vector with independent basis vectors, $\vec p = \sum_j u_j\vec u_j$. 

As a numerical procedure, each iteration may be more expensive than other numerical methods, since it involves calculation of the eigenvalues and eigenvectors. However, the promise of using gradient ascent is to reduce the number of iterations required to reach the solution. For lower dimension problems, the first order derivatives and even second order derivatives can be worked out by hand, and input into the algorithm to use. In terms of physics, the computation of changes in eigenvalue and entropy gives us a measure of understanding about the optimization problem. For example, we can recognize the eigenvalue perturbation formula from quantum mechanics --- therefore we can make use of familiar second order perturbations of eigenvalues if we need to implement a Newton optimization algorithm. In practice, the computation of the second derivatives are a lot of more expensive numerically, to be worth the reduction in number of iteration steps.

That means, if we have an objective function $f$, then we will instead work with $g$ defined as
\begin{equation}
  g = \left\{ \begin{array}{l l}
  \lambda_\text{min}\,, & \quad \lambda_\text{min} < 0\,,\\
  f\,, & \quad \lambda_\text{min} \geq 0\,. \end{array}
 \right.
\end{equation}
The rationale for doing so is such that we can start our optimization anywhere within the probability simplex. At any iterate, we test if the minimum eigenvalue is negative. If yes, we will calculate the gradient of the minimum eigenvalue, $\nabla\lambda_\text{min}$, and move the next iterate to a point with greater value for the minimum eigenvalue. In other words, we always perform maximization at any point, no matter whether the function defined on that point is the minimum eigenvalue or the objective function. Eventually, this guides iterates into the permissible region, and we can maximize $f$ properly.

\subsection{Random sampling}
We also require some numerical methods for random sampling of states. In general, there a few ways to `randomly' pick a state, and these ways are connected to how the state is prepared and/or how the space of states is measured. Without going into the mathematical intricacies or the endless debate over which measure might be better, we will return to a more physical approach to treating this problem. The problem of sampling a state, and its connected measure is settled by considering tomography as `game' played between two persons. One person, perhaps Alice, has prepared the state is some manner unknown to us, and the other, perhaps Bob, is now trying to guess it. The setup is just like a game of \emph{Mastermind}. We may treat Alice as randomly picking a state, but according to the specifics of how she prepares the state. Some ways of preparing a (theoretical) qutrit could be:
\begin{enumerate}
  \item Take 3 orthogonal pure states uniformly according to the Haar measure, and mix them, picking uniformly from the probability simplex to perform this mixture.
  \item Take the MUB set, and pick uniformly from the probability simplex to get a random $\rho$.
  \item Take a pure state at random (uniformly according to Haar measure), and mix with the completely mixed state.
  \item Take $K$ pure states (uniformly according to Haar measure), and mix them in some manner. 
\end{enumerate}
For different methods, the corresponding eigenvalue distributions, purity distribution, or distribution for some function of relevance will all be different. Since Bob does not know which of these preparation methods might be used, he should consider as many of them as possible. 

Additionally, we need to randomly sample states from the permissible region in some manner. The most basic way to do so would be Monte Carlo style, uniformly sampling the probability simplex, and discarding points corresponding to density matrices with negative eigenvalue. This can be done quite rapidly with Cholesky factorization. The only problem is that some of the regions could be small compared to the simplex. With random sampling in this manner, it is quite easy to miss the region altogether. When this happens, it is recommended to switch back to maximizing the minimum eigenvalue in order to have a better chance at locating the region. 

A possible way to improve this is the technique of importance sampling, using a handful of trial random points to roughly locate the region, and then using a Gaussian distribution with covariance matrix computed from the trial points to sample at greater detail. However, this method is more expensive to compute since one needs to calculate the p.d.f of the Gaussian, and it still suffers when the trial points are \emph{unable} to find the region.

\section{Maximum von Neumann entropy estimator, MvNE}
\begin{figure}
  \centering
  \includegraphics[width=0.7\textwidth]{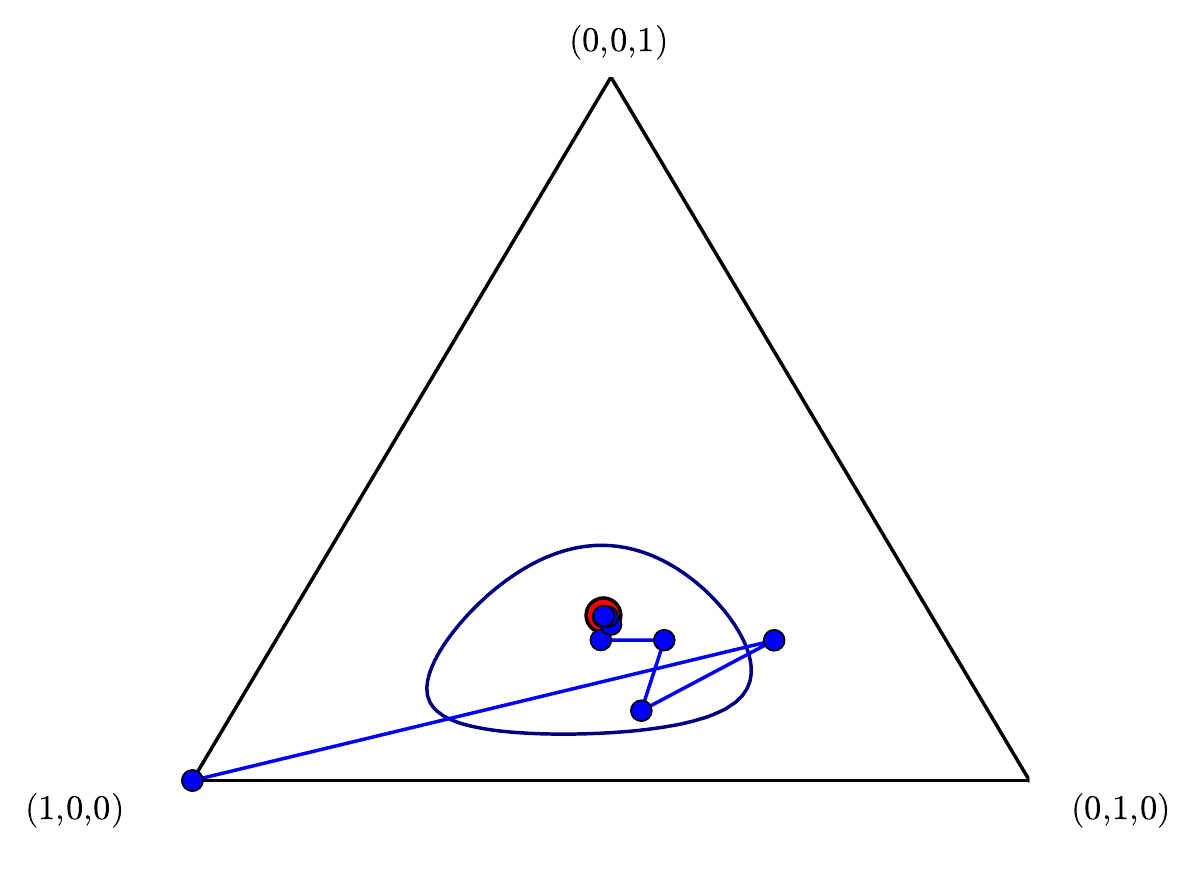}
  
  \caption{Maximizing von Neumann entropy with a piecewise function and gradient ascent. Note how it converges onto the red point computed independently by another numerical method, starting from one corner of the simplex.}
  \label{fig-qent1}
\end{figure}

\sectionmark{MvNE}
Numerically, the state with maximum von Neumann entropy lies entirely within the permissible region, so in order to find it we simply define the piecewise function as stated in the previous section, using von Neumann entropy as the objective function and maximize (figure \ref{fig-qent1}). 

The variation of von Neumann entropy with respect to vector coefficients $c_j$ is worked out in a few steps as follows,
\begin{align}
	\delta S = - \Tr{\delta\rho \log\rho + \delta\rho} = -\Tr{\delta\rho\log\rho}
\end{align}
since the variation of $\rho$ only involves traceless matrices, as easily seen from the parameterization of $\rho$.
\begin{align}
	\delta S = -\Tr{\log\rho\cdot\left(\sum_j\delta c_j\Lambda_j\right)}= - \sum_j \Tr{\log\rho\cdot\Lambda_j}\delta c_j.
\end{align}
Now by considering $\log\rho = \sum_i \ketbra{E_i}{E_i}\log\lambda_i$, where $\lambda_i$ and $\ket{E_i}$ are the eigenvalues and eigenvectors respectively of $\rho$, then we work out the variation of entropy to be
\begin{equation}
	\delta S = -\sum_j \sum_k \log\lambda_k\,\Tr{\ketbra{E_k}{E_k}\Lambda_j}\delta c_j\,.
\end{equation}
Therefore, since we are already calculating the eigenvalues and eigenvectors of $\rho$ in order to perform gradient ascent on entropy, we might as well make use of the minimum eigenvalue to check whether our iterates are feasible, and use its gradient to move towards the permissible region.

On the simplex of eigenvalues, the von Neumann entropy is a concave function with maximum at the completely mixed state. But more importantly, it is concave also on the permissible region $\Gamma$ with its maximum in interior of the region. If one tries instead to maximize the minimum eigenvalue to find the MvNE, reasoning that the minimum eigenvalue is as large as possible for the MvNE under the constraints, one will find that this method does not work.
\begin{figure}
	\includegraphics[width=0.7\textwidth]{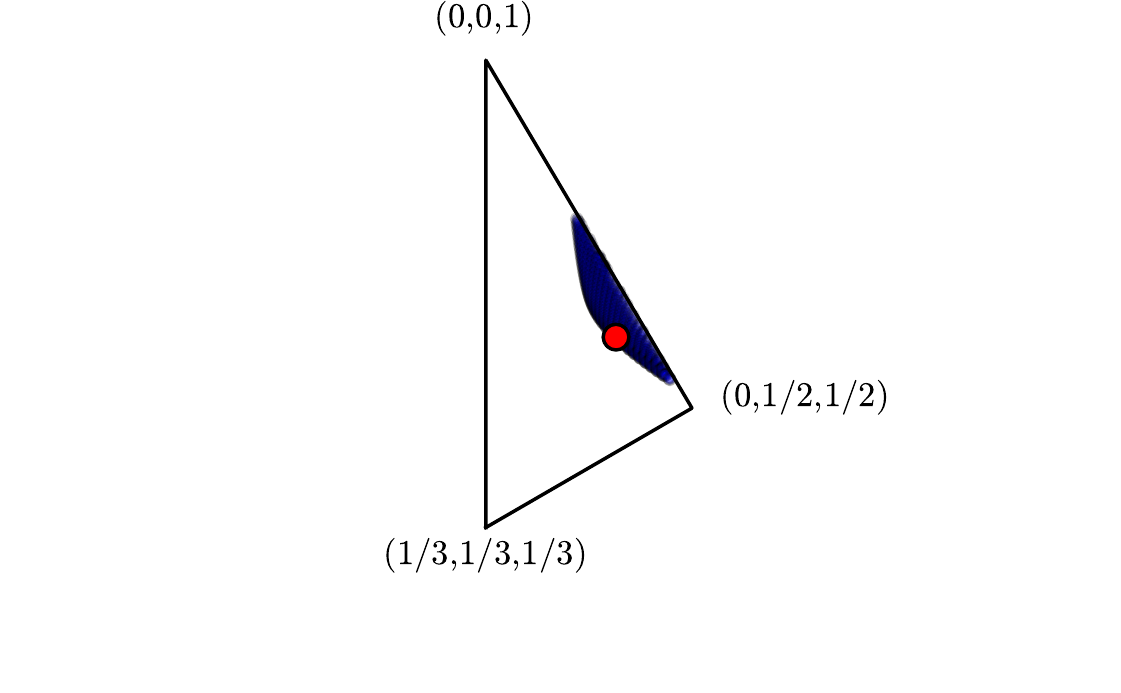}
	\caption{Maxizing von Neumann entropy in the space of eigenvalues under the same constraints as described. The red spot marks the MvNE, and it lies on the boundary of possible eigenvalues. The simplex is smaller, due to reason that the eigenvalues are equivalent under permutation.}
	\label{fig-eigenoptim}
\end{figure}
From figure \ref{fig-eigenoptim}, the MvNE has eigenvalues that lie on the boundary due to constraints in the eigenvalue simplex. If we simply maximize the minimum eigenvalue, we have to take care of the constraints, which is not easily expressed in the eigenvalue space.

\section{Maximum Shannon entropy estimator, MSE}
\begin{figure}
  \centering
  \includegraphics[width=0.7\textwidth]{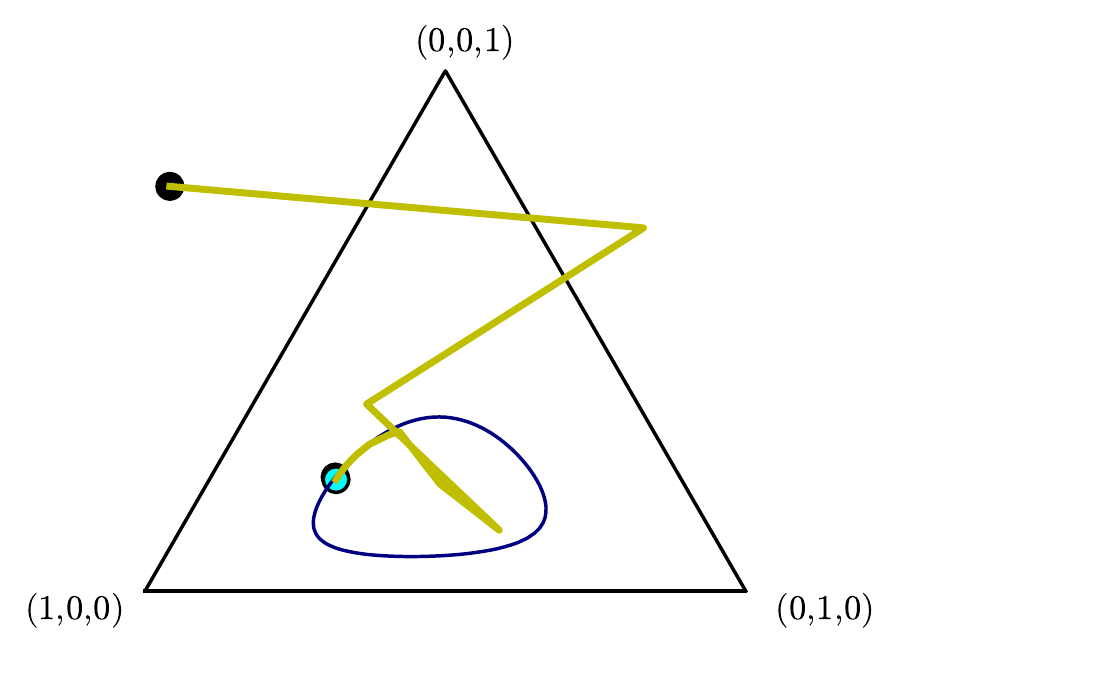}
  \caption{Implementation of the log barrier method to solve for the state with maximum Shannon entropy for a particular future measurement that is \emph{not} the computational basis.}
  \label{fig-logbarrier2}
\end{figure}

\sectionmark{MSE}
The computation of the maximum Shannon entropy estimator (MSE) is a little more tricky than for von Neumann entropy. This is because Shannon entropy is defined on the entire probability simplex, and the unconstrained maximum is where the probability for the outcomes is uniform. However, since the permissible region is only a subset in the probability space, the state with maximum entropy could lie on the boundary if the unconstrained maximum lies outside the permissible region. Therefore, in order to continue to use gradient ascent, we need to `enhance' the objective function with the log-barrier method \cite{boyd2004convex}. The spirit of this method is that the iterates are guided into the permissible region initially, through maximizing the minimum eigenvalue. Once the iterates become feasible, the iterates are not allowed to leave the permissible region, as they will encounter a logarithmic barrier at the boundary. This means that if the constrained optimum lies on the boundary, the barrier allows the iterates to converge close to the optimum safely without the iterates becoming infeasible (figure \ref{fig-logbarrier2}). The algorithm starts by first checking the state corresponding to uniform probability values. If this point is within the permissible region, we have found the MSE and can terminate. If not, we can use this point as the starting iterate and begin optimization.

The function that we will use is
\begin{equation}
  g = \left\{ \begin{array}{l l}
  \lambda_\text{min}\,, & \quad \lambda_\text{min} < 0\\
  -\sum_i p_i\log p_i +\frac{1}{t} \log\lambda_\text{min}\,, & \quad \lambda_\text{min} \geq 0 \end{array}
 \right.
\end{equation}
where $t$ is a freely adjustable parameter controlling the strength of the log barrier. A value of $t=1e-4$ or $1e-5$ is sufficient. If one desires to produce a sequence of sub-optimal points that eventually converge upon the true optimum, then one can repeat the optimization for a sequence of decreasing $t$ values, starting perhaps at $t=1e-2$, using the terminating iterate for optimization with the current $t$ value to be the starting iterate of the next optimization run with the next $t$ value. The theoretical details of this can be found in \cite{boyd2004convex}.

After the iterates have been guided into the permissible region, optimization can proceed by use of Newton method. The Hessian of entropy is easily computed. On the other hand, the Hessian of the minimum eigenvalue, involding second order perturbations of the eigenvalues more expensive to compute. Instead of the Hessian of the minimum eigenvalue, we can use the dyadic of the gradient $(\nabla \lambda_\text{min})(\nabla \lambda_\text{min})^T$ instead. Optimization terminates when the expectation of the total Hessian matrix with respect to optimization direction $\vec s$, $\vec s^T\cdot\text{Hess}\cdot\vec s$, is sufficiently small.

\subsection{Different future measurements}
The definition of Shannon entropy for quantum systems involve choosing a measurement basis. And as we mentioned in chapter 1, a set of $\Lambda_j$ with 8 non-zero singular values in the matrix $[M]_{jk}\equiv\Tr{\Lambda_j\Lambda_k}$ allows us to reconstruct the qutrit fully. Therefore, if we have one basis unmeasured, then choosing a measurement and maximizing Shannon entropy gives us the MSE.

Different physicists choosing to use different future measurement have different parameterizations of the $\rho$ and sets up different probability simplexes to measure. The change of the coordinates from one observer to another is related by an affine transformation, as is evident by looking at equation \eqref{eq-probabilityreconstruct}. 
Consider now Alice who chooses to use set $\{\Lambda_j\}$ to parameterize a state $\rho$, compared to Bob who chooses to use set $\{\Lambda'_j\}$ to parameterize the same state. In their chosen sets, there are common elements, since the measured bases are the same for both, and are therefore fixed. What is freely chosen are the unmeasured bases. Accordingly, for the same state $\rho$, Bob's probability values for measurement outcomes are going to be different compared to Alice's,
\begin{align}
	\vec{p}\,' - \overrightarrow{\frac{1}{3}} = C M^+\left(\vec{p} - \overrightarrow{\frac{1}{3}}\right)\,,
\end{align}
where $C$ is the change of basis matrix with elements given by $\Tr{\Lambda'_j\Lambda_k}$.
Then we can separate out the unmeasured components from the measured ones in this matrix equation. 

\begin{figure}
  \centering
  \includegraphics[width=0.7\textwidth]{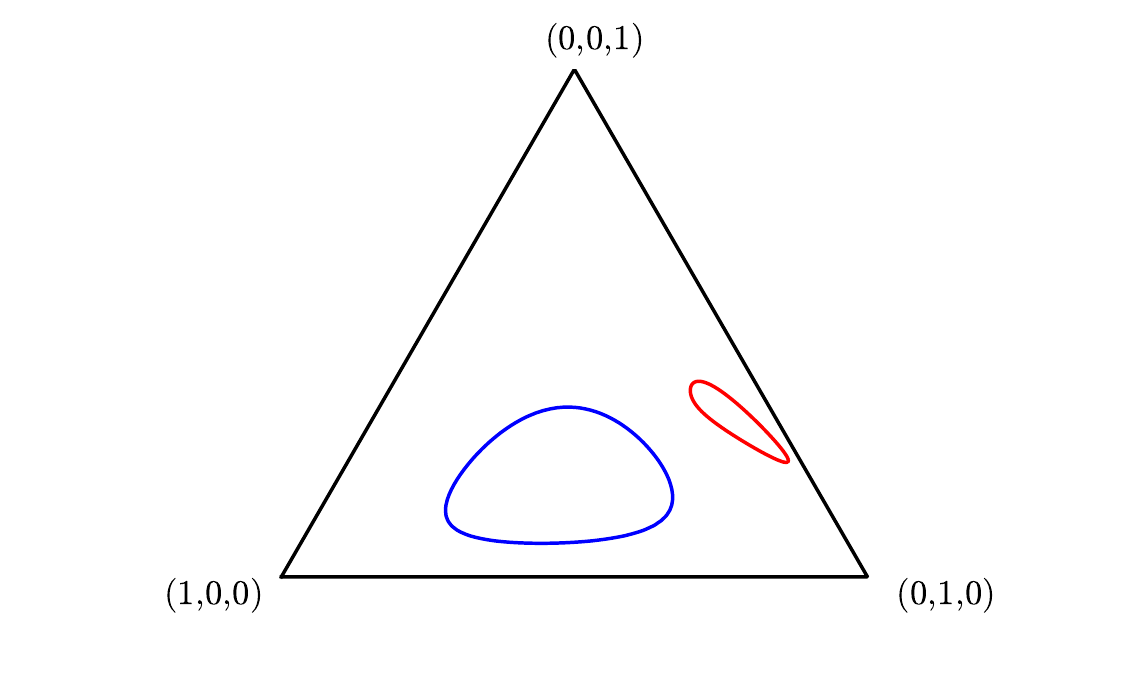}
  \caption{For two different measurements, the permissible region undergoes an affine transformation that changes its shape, area, and location. Here two probability simplexes are superimposed on top of one another, to show the relative sizes and positions of the permissible regions describing the same state. That means an Alice using some future measurement might set up her simplex and see the blue region, while Bob using some different measurement would set up his simplex and see the red region.}
  \label{fig-probtransform2}
\end{figure}

Overall, Alice's future (unmeasured) probabilities $\vec f$ and Bob's future (unmeasured) probabilities $\vec q$ are related by a linear equation of the form
\begin{equation}
  \label{eq-probtransform}
  \vec f = V\vec q + \vec\beta\,,
\end{equation}
where $V$ depends on some unitary matrix that transforms Bob's future measurement basis to Alice's future measurement basis, and $\vec\beta$ is a column of coefficients that depend on measured probabilities of $\rho$. This equation describes an affine transformation between two observers using different future measurements (figure \ref{fig-probtransform2}).
In this transformation in general, some probabilities are no longer probabilities in another observer's point of view (they become negative), but the permissible region always remain possible under transformation, something that is consistent with quantum mechanics. Otherwise, some measurement bases would not be observable. 

A point of note is that if Bob chooses some measurement that is a linear combination of already measured bases, he will find that his permissible region shrinks to become a point, since $\mathrm{det}\,V=0$ in \eqref{eq-probtransform} and the relation is no longer bijective. This is simply an indication to Bob that his chosen measurement does not give him any additional information about the system under measurement.

\subsection{Average over ensemble of physicists}
\begin{figure}
  \centering
  \includegraphics{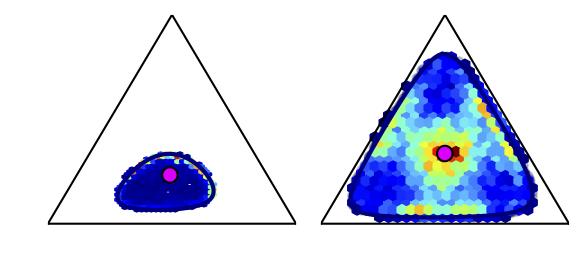}
  \caption{By solving for the MSE for many different future measurements, one obtains a distribution. Clearly, this distribution depends on how the measurements are sampled, and here they are sampled uniformly from Haar measure. The two simplexes show two examples for different prior information (different off-diagonal matrix elements), described in the same computational basis, hence with different permissible regions. Pink point plots the mean MSE.}
  \label{fig-logbarrier}
\end{figure}

If you are either Alice or Bob, you might well be justified in insisting on maximizing Shannon entropy using your chosen measurement. And this is fine, since there is no \emph{a priori} reason as yet to indicate that one measurement might be more privileged than others. But however if we have yet to decide on which future measurement, it might be better to consider estimators that are non-informative about information we do not yet have. One compromise to this is to consider an ensemble of physicists, all having chosen a different measurement and maximized Shannon entropy, and choose the average state from the ensemble of maximum Shannon entropy estimators as \emph{the} estimator. As we have mentioned previously, this entails the solution of many optimization problems in order to form a distribution, but the result is illustrated in figure \ref{fig-logbarrier}. Formally, we mean that we take the estimator to be
\begin{equation}
	\hat\rho = \int_{SU(d)}(\D\, \phi(U))\,\operatorname*{arg\,max}_\Gamma\, H(\rho,U\{\Pi_j\}U^\dagger)
\end{equation}
where we recognize that that the columns of unitary matrices are orthonormal bases, and $(\D\,\phi(U))$ is a integration measure which in this case we can take to be the Haar measure, but other ways are possible also. In some sense, the Haar measure is a `uniform' measure on the space of measurement bases, so there is some justification for taking the mean estimator in this manner. Of course, the question now is whether there can be a better distribution that we can assign to the space of measurement bases, so we have simply changed the problem of assigning a `best state' to now one of assigning a `best measurement'.
What was observed is that many of the MSE have a tendency to be on the boundary. But if prior information about the system results in a permissible region that is large compared with the probability simplex, then many MSE also tend to be near the uniform distribution for the observable outcomes of the measurement, that is near the center of the simplex. 

It is also observed that other than the unsavory idea that we now need to introduce some way of averaging over measurements, the resulting average estimator (the pink point plotted in figure \ref{fig-logbarrier}) is almost identical to the MvNE. Overall, we have found this estimator to be disadvantageous to use compared to other methods, due to its large computational costs.

\section{Bayesian mean estimators}
We may instead choose to assign a noninformative prior over relevant parameters, and take the mean to form the Bayesian mean estimator. By noninformative, we mean uniform distribution over the parameters if applicable, or Jeffreys prior if the distribution needs to be invariant under several equally applicable parameterizations.

\subsection{Jeffreys prior?}
If we are to do as Jeffreys original solution to the problem, and consider a prior that is invariant under all possible parameterizations, we will end up with a distribution that is proportional to $\frac{1}{\sqrt{p_1 p_2 p_3}}$. However, we should question whether such a consideration is applicable for our problem. We have seen that the permissible region is a closed, finite region. Under Jeffreys' original solution, all monotone transformations of parameters are considered. Some of parameterization might not be physically relevant to the problem. A more careful consideration involves the thinking about the physically relevant parameters, and how they are measured in the laboratory. For our problem at hand, the relevant parameters are the permissible region of states $\Gamma$, and the unmeasured bases $\{\Pi_j\}$. In the laboratory, we measure frequencies of measurement outcomes. Because the assignment of prior is for the purpose of using Bayes rule, and how we use Bayes rule is closely tied to how we receive information about the system, that is, through measurement. Therefore, by viewing the probability of measurement outcomes as coordinates, and the measurement bases as coordinate basis for the system, we assign a distribution that expresses the fact that we do not favor any particular state. That means we first try to assign the uniform distribution on $\Gamma$, and then check if this distribution is consistent under other physically equivalent situations. 

One conceptual difficulty of using the probability distribution $\frac{1}{\sqrt{p_1 p_2 p_3}}$ as the prior is that when one does this, one is really assigning a distribution to the probability simplex, without regard to the underlying physics. Consider more closely what we mean when we use this distribution as the prior. Since this distribution is independent of whether the system is classical or quantum (it ignores the permissible region), we are really assigning a prior that for a system that is classical. That is, the scenario we are dealing is really: Suppose I walk into a laboratory and see a black box that is outputting random outcomes like ``1,3,2,2,1,3,3,2,1,2,3,1,2,3,$\cdots$". Without more knowledge about this black box, I will assume that it is a classical random system, and so the prior on $p_1, p_2, p_3$ for the outcomes is defined on the whole simplex. Being the good Bayesian that I am, I use the non-informative prior $\frac{1}{\sqrt{p_1 p_2 p_3}}$. But then a colleague of mine, who is working on the experiment, walks in at this moment, and tells me casually: ``Oh, I am measuring a qutrit. In fact I have already measured it in these bases," and proceeds to tell me those measurement data. Since we know that the possible probabilities are constrained by the data, do we now enforce this constraint through the step function $\eta(\Gamma)$? Bayes rule would say, posterior is proportional to likelihood times prior, and so we find that our posterior is now $\frac{1}{\sqrt{p_1 p_2 p_3}}\eta(\Gamma)$. However, we are now committing an error: Originally, the $p_i$ are referring to a classical random system. But now we are using them in vector coefficients for a quantum system. In particular, there is no uniqueness to the posterior, as another person who uses a different parameterization would end up with another posterior $\frac{1}{\sqrt{p_1 p_2 p_3}}\eta(\Gamma')$, which would assign a different probability distribution to the same set of states, since the Jeffreys prior is not invariant under change of choice for future measurement basis\footnote{We will see why subsequently.}. Though we have stated the statistical operator to be a subjective concept, dependent on the knowledge of observers, we nonetheless believe that two observers given the same amount of information would produce the same prediction if both of them maximized their given knowledge individually. This is simply a consistency postulate\footnote{Much like the rational choice hypothesis in economics}.

\subsection{Invariant under choice of future measurements}
Due to the puzzle of possible different measurements to complete the tomography of the system, we should examine whether the uniform prior on $\Gamma$ or the Jeffreys prior for probabilities remains invariant under such reparameterization. Again, we must reiterate against assigning a prior on the entire probability simplex and then enforcing the permissible region constraint through the step function. Suppose we can assign probability distribution $f$ to the probability simplex $\mathrm{prob}_\Omega$ for chosen measurement basis $\Omega$. Let $T$ be the transformation between $\Omega$ and $\Omega'$.
However, if we have $f:\mathrm{prob}_\Omega \rightarrow \mathbb R$, then this function cannot be bijective. This is because under transformation $T[f]$, some probabilities represented in $\mathrm{prob}_\Omega$ would not be represented in $\mathrm{prob}_{\Omega'}$. We have seen that in general, affine transformation of probabilities might not be probabilities. However, probabilities within the permissible region remain probabilities under transformation. Therefore, we can only define our function $f:\Gamma\rightarrow \mathbb R$ on the domain of the permissible region $\Gamma$.

Now suppose Alice uses measurement $\Omega$, and Bob uses measurement $\Omega'$. Since the two of them are independent of each other, we might imagine Alice assigns her prior $f$ independently of Bob's prior $g$. If Alice wants to calculate the prior probability that some $\gamma \subseteq \Gamma$ contains the true state, she writes
\begin{equation}
  \label{eq-aliceInvariant}
  \mathrm{prob}(\gamma:\gamma\subseteq\Gamma) = \int_\gamma (\D \vec p)\,f(\vec p)\,.
\end{equation}
Likewise, Bob does the same for his prior $g$. Now since Alice and Bob are describing the same state, they have to be consistent. That means,
\begin{align}
  \label{eq-consistentprob}
  \mathrm{prob}(\gamma:\gamma\subseteq\Gamma) &= \mathrm{prob}(T[\gamma]:T[\gamma]\subseteq T[\Gamma])\,,\nonumber \\
  \int_\gamma (\D \vec p)\,f(\vec p) &= \int_{T[\gamma]} (\D \vec p')\, g(\vec p') = \int_{\gamma} (\D \vec p)\,J\,g(\vec p)\,,
\end{align}
where $J$ is the Jacobian determinant --- a constant in this case for an affine transformation. Written out more fully, we have the condition that,
\begin{equation}
  \label{eq-consistentprob2}
  f(\vec p) = J \,g(V^{-1}(\vec p - \vec\beta))\,,
\end{equation}
where we have written out the transformation of $\vec p$ more explicitly. As a reminder, the vector $\vec p$ refers to unmeasured probabilities only. 

If $f,g$ are uniform distributions, equation \eqref{eq-consistentprob2} holds, up to a inconsequential normalization factor. On the other hand, due to the translation factor $\beta$, the Jeffreys prior of the previous subsection is not invariant under the required affine transformations. Numerical trials show that in general that Alice may find that $\Gamma$ appears larger relative to the simplex than for Bob, or vice versa, on top of location changes within the simplex. In other words, though the both Alice and Bob may choose their measurements $\Omega$, $\Omega'$ with equal right to complete the reconstruction of state, the constraints appear differently for them in terms of the size of the permissible region and its location within the simplex. It is a hint that perhaps things are different for Alice and Bob.

\subsection{Center of mass estimator, COM}
Using the uniform distribution defined on the simplex, we can define the mean as the as center of mass estimator, COM. As the name implies, it is simply the centre of mass of the region as if it had uniform mass density. Intuitively, this estimator is the same in all future measurement bases, since affine transformations do not change the position of the center of mass relative to the mass.

Numerically, it can be found in a simple way by sampling the simplex in a uniform way and accepting only points which lie in the permissible region. The average of these points is the COM. 

\section{Random estimator}
The reader may well be of the opinion that since one does not know which state in the permissible region might be the true state, then instead of using the center of mass estimator, he may choose instead to simply pick any state at random from the permissible region as the estimator. The sampling would also be based on the uniform distribution that we have proposed in the previous section. We state not to promote the use of this estimator, but to set up another estimator as comparison. The random estimator has the similar flavor to a student that has given up in the exam. Faced with no knowledge of which choice might be in the correct answer in the multiple choice questions, he/she simply picks one at random. 

\begin{savequote}[200pt]
	All sodium salts burn yellow. However, a piece of pure ice held in colorless flame confirm the assertion that ``whatever does not burn yellow is not sodium salt". Does this mean that the observation of the flame for pure ice constitute as evidence for the statement about all sodium salts burning yellow?
	\qauthor{C. G. Hempel, formulating the Hempel's paradox}
\end{savequote}
\chapter{Comparison of estimators}
Here we will compare the various estimators. We can do this by randomly sampling a true state, measure the probability values for some of bases, and applying the various estimator strategies to `fill in the blanks' for the rest of the unmeasured bases. We remind the reader that the MvNE and COM is independent of the future bases that are going to be measured, while the MSE and the random estimator is dependent.

\begin{figure}
  \centering
  \includegraphics[width=0.7\textwidth]{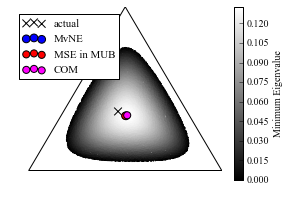}
  \caption{The MvNE, MSE, COM for a randomly chosen mixed state. The true state used to generate the permissible region is marked with an `x'.} 
  \label{fig-estimatorcompare4}
\end{figure}

\begin{figure}
	\centering
	\includegraphics[width=0.7\textwidth]{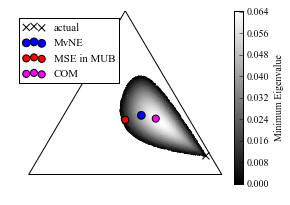}
	\caption{The MvNE, MSE, COM, for a randomly chosen pure state. The true state used to generate the permissible region is marked with an `x'.}
	\label{fig-estimatorcompare3}
\end{figure}

\begin{figure}
	\centering
	\includegraphics[width=0.7\textwidth]{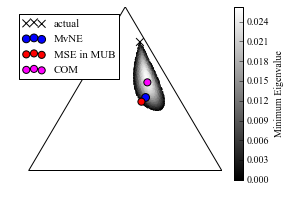}
	\caption{The MvNE, MSE, COM for a randomly chosen pure state. The true state used to generate the permissible region is marked with an `x'.}
	\label{fig-estimatorcompare5}
\end{figure}

\begin{figure}
	\centering
	\includegraphics[width=0.7\textwidth]{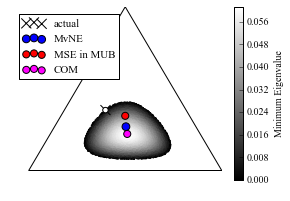}
	\caption{The MvNE, MSE, COM for a randomly chosen rank 2 state. The true state used to generate the permissible region is marked with an `x'.}
	\label{fig-estimatorcompare7}
\end{figure}

It was observed that for highly mixed states, the permissible region associated has large area compared with the simplex, and subsequently many estimators have a tendency to be close to the centre of the simplex, an example of which can be seen in figure \ref{fig-estimatorcompare4}.

If the true state were a pure state, then the boundary of the permissible region will have a `kink' or cusp, where the pure state would be located (figures \ref{fig-estimatorcompare5}, \ref{fig-estimatorcompare4}). While this seem like an advantageous piece of information, we must remember that there is no easy way to check if the boundary has such a extremal point. In the appendix, there is a short piece of algorithm to check boundary states of the permissible region, but it is still rather resource consuming. Moreover, given prior data, if we find that the permissible region possesses such an extremal point, it does not guarantee that the true state is a pure state. We have only found that this a likely pattern through numerical experimentation. More analysis is required to confirm this. We have also observed that if the true state were a pure state, the permissible region is frequently likely to be small (figure \ref{fig-estimatorcompare5}). In fact most of the times while simulating data from a pure state, the permissible region is so small that it cannnot be plotted on the simplex, and the algorithms that search for the estimator frequently fail as well. In these cases, it is much better to approximate the likely location of the estimator by maximizing the minimum eigenvalue as much as possible. The resulting estimator will likely give clues to the identiy of the pure state.

If the true state was a rank 2 state, meaning that it is still rank deficient, then the true state will sit on the boundary of the permissible region (figure \ref{fig-estimatorcompare7}). In these cases, the permissible region is neither large like those belonging to highly mixed states, nor small like those belonging to pure states. But in these cases, since the true state lies on the boundary, the center of mass estimator which is at the center of the permissible region (geometrically speaking), will not be as close to the true state as compared to the maximum entropy estimators. Particularly, we should find that the MSE has a chance of being much closer to the true state.

\section{Estimator performance}
One way to test how the various estimators perform is to consider how close the estimators are to true states. The distribution due to sampling of true states influences the distribution of distances between true state and estimator. Ideally, a good estimator would have such a distribution with mean as close to the zero distance, and variance as small as possible. Consider if we have perfect knowledge of the system. Then the distribution of estimator to true state distances would have mean 0 and infinitely small variance, like a Dirac delta function. If the mean distance is non-zero, then there is either lack of knowledge about the system such that we are unable to pin down its system which manifests as random variations of the estimator, or that there is a persistant bias or error in the estimation method \cite{lehmann1998theory}.

In the figures that follow, they are produced by repeatedly sampling true states (100000 times), calculating the probability for measurement outcomes of a few measurement bases, and using this as prior information in performing various estimator strategies. Distance between the true states and the resulting estimators are calculated, and a distribution of such distances are plotted. There are also cases when the numerical method fails to compute estimators. When this happens for any particular method, `not-a-number' is recorded for all the methods. However, these pathological situations are rare ($\sim$3\% of the sampling rate), only occurring when the permissible region is particularly small, the only fault lying with the numerical method for not being sufficiently fine-tuned.

Two queries are immediate. One concerns how the true states are sampled. Since we do not know in reality how true states might be sampled, we will present results for a few sampling methods. Instead of thinking of \emph{the} sampling method, we can think of sampling true states as a part of game between Charlie and us. Charlie is in the laboratory and he mixes up states by some manner, and it is always in a fully physical manner. For example, he could prepare 3 orthogonal pure states, and mix them in some ratio. Or he could take a pure state and mix with the completely mixed state. The situation is similar to Bertrand's paradox; in these situations, we can either try to compute an answer that is independent of the unspecified information, or we can say, `It depends on who is doing what.' Since we have no particular need to find out exactly what is it that Charlie used to produce the quantum systems, we simply test a number of scenarios. The second query is related to the first one; it concerns how distances are measured. We will simply test a few distance measures at this stage. 

We tested true state samples based on uniformity in Hilbert-Schmidt distance. This method of sampling was found to be identical to sampling uniformly from a probability simplex, using a qutrit SIC-POM (9 outcomes) to reconstruct the state, and also identical to using the MUB set for reconstruction, using an appropriate number of independent probability simplexes with 3 outcomes. We also tested sampling by taking 3 random orthogonal states (taken from unitary matrix random on Haar measure), and selecting eigenvalues uniformly from the eigenvalue simplex. Lastly, we also sampled true states based on mixing a randomly selected pure state and the completely mixed state. 

Without a clear reason to favor a particular distance measure at this stage, a few more commonly used distance measures between states were tested. They are the Hilbert-Schmidt distance defined on $[0,1]$, zero only if two states are the same; fidelity on $[0,1]$, one only if two states are the same; and quantum relative entropy on $[0,1]$, zero only if two states have the same set of eigenvalues. The use of Hilbert-Schmidt distance was motivated in that it behaves intuitively much like the Euclidean distance between matrices; while for fidelity, the motivation was that it is a similarity measure on its own, but is also connected to other proper distance measures such as the Bures distance. The use of quantum relative entropy was primarily motivated through the fact it is measure of divergence of statistical mixtures of orthogonal states. It is similar to the use of relative entropy in classical probability theory. It was thought that perhaps the maximum entropy estimators might perform especially well on this measure.
\begin{figure}
  \centering
  \includegraphics[width=0.7\textwidth]{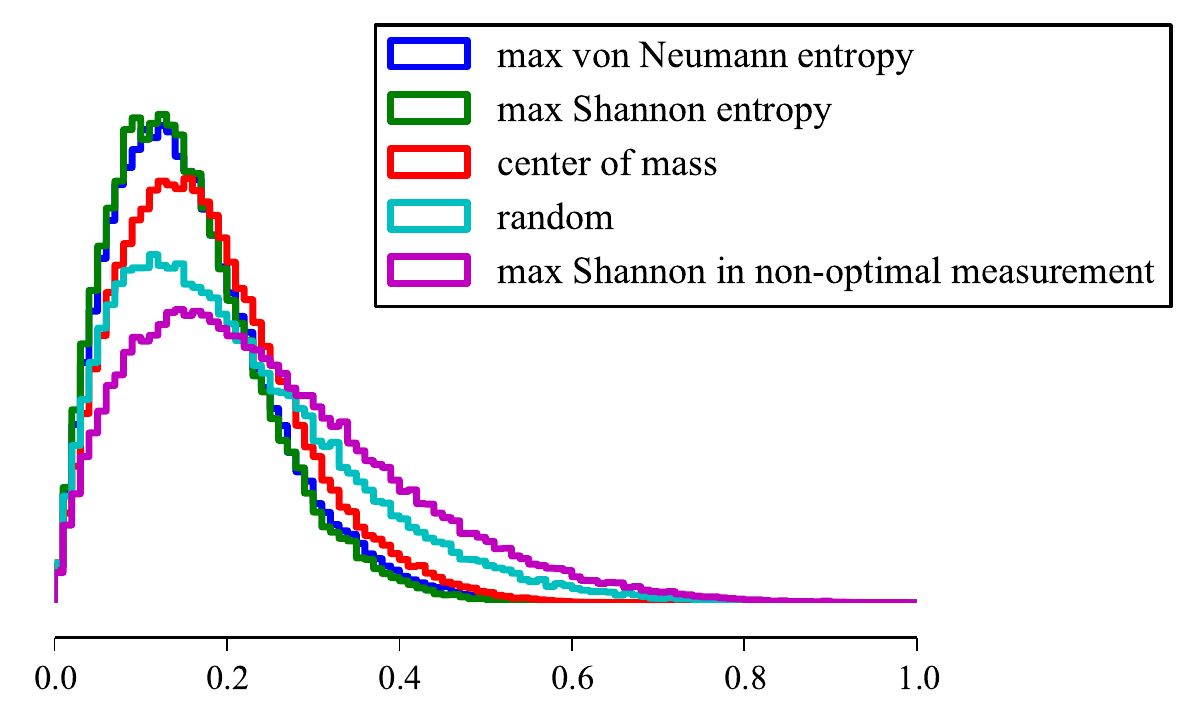}
  \caption{Sampling uniform based on Hilbert-Schmidt distance, distance measured with Hilbert-Schmidt distance. The maximum von Neumann entropy, maximum Shannon entropy with unmeasured MUB, and center of mass estimators are quite competitive, in contrast to the random estimator and maximum Shannon entropy in non-optimal measurement.}
  \label{fig-estimatorcompare3HS}
\end{figure}

\begin{figure}
  \centering
  \includegraphics[width=0.7\textwidth]{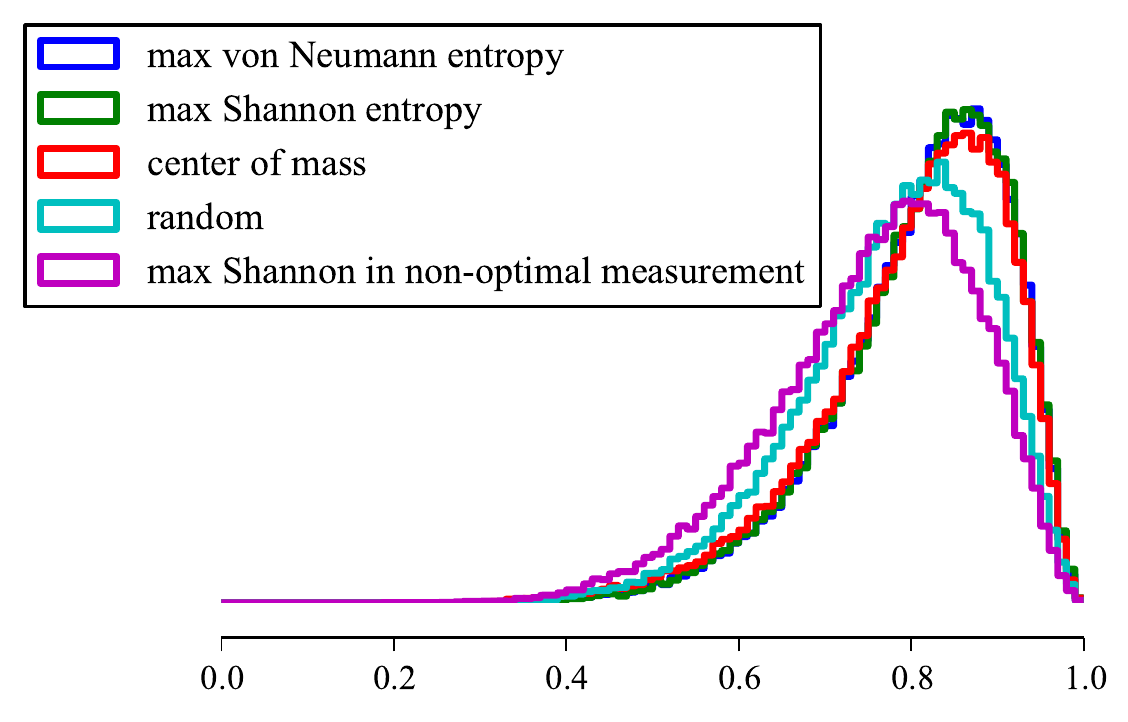}
  \caption{Sampling uniform based on Hilbert-Schmidt distance, distance measured with fidelity.}
  \label{fig-estimatorcompare3Fid}
\end{figure}

\begin{figure}
  \centering
  \includegraphics[width=0.7\textwidth]{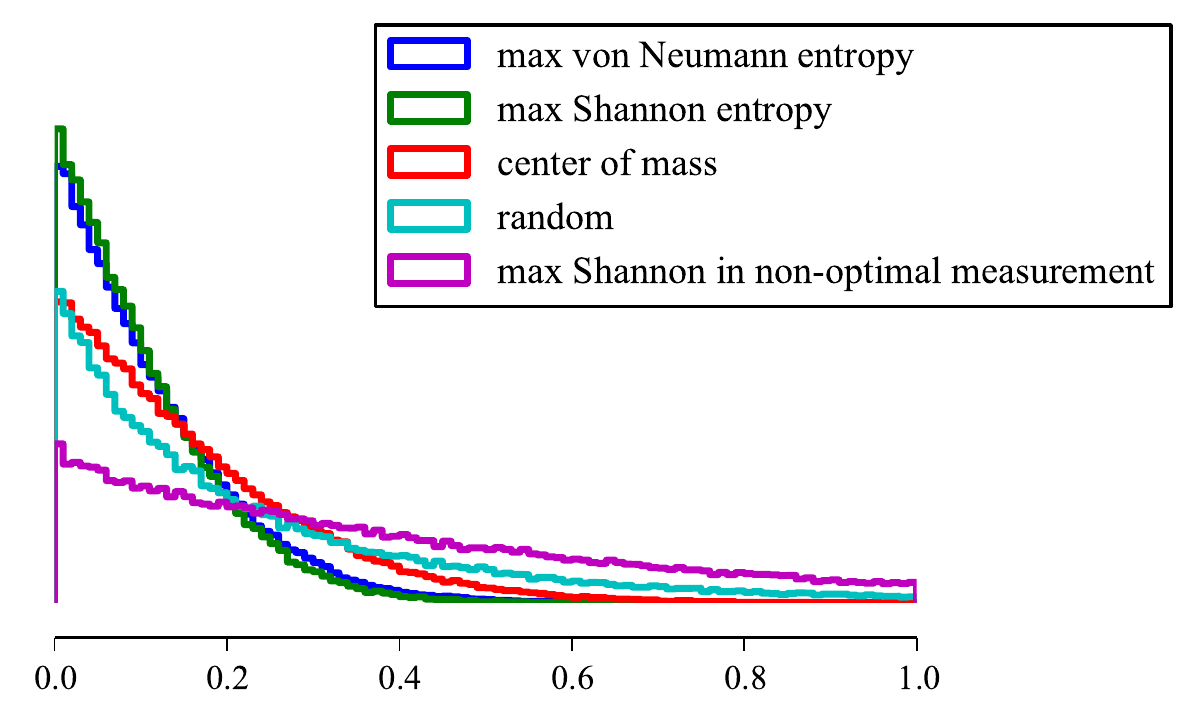}
  \caption{Sampling uniform based on Hilbert-Schmidt distance, distance measured with quantum relative entropy.}
  \label{fig-estimatorcompare3RelEnt}
\end{figure}

\begin{figure}
  \centering
  \includegraphics[width=0.7\textwidth]{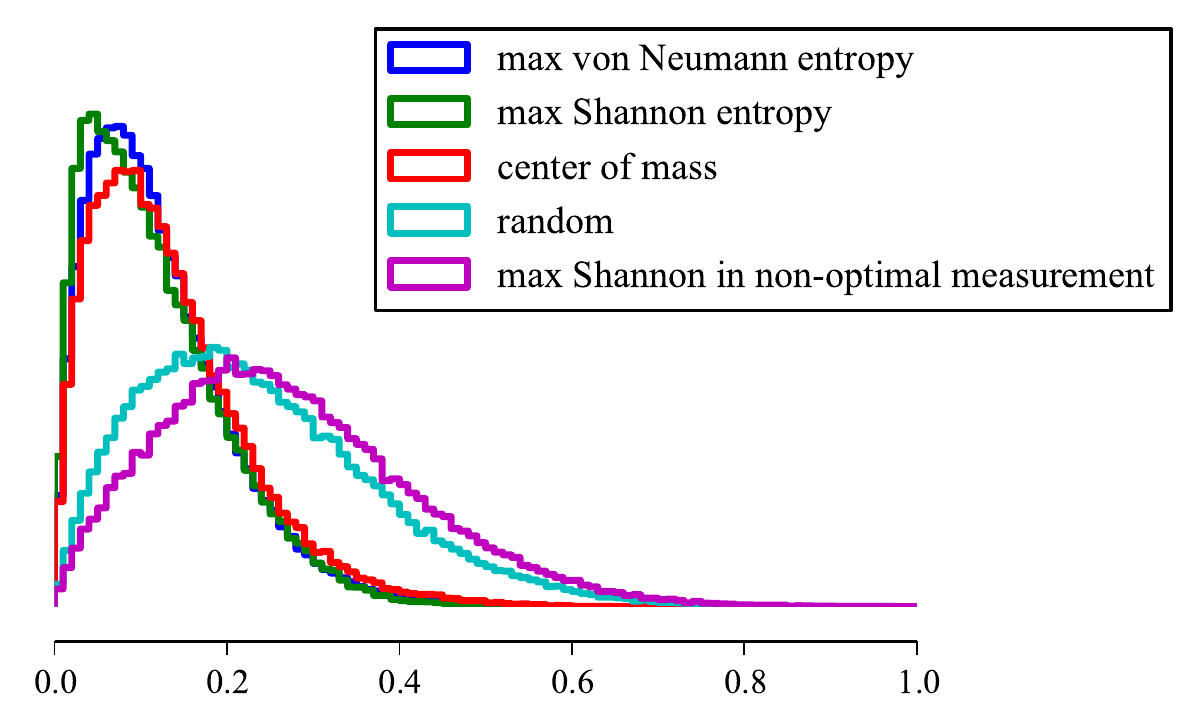}
  \caption{Sampling uniform on eigenvalue simplex, distance measured with Hilbert-Schmidt distance.}
  \label{fig-estimatorcompare4HS}
\end{figure}

\begin{figure}
  \centering
  \includegraphics[width=0.7\textwidth]{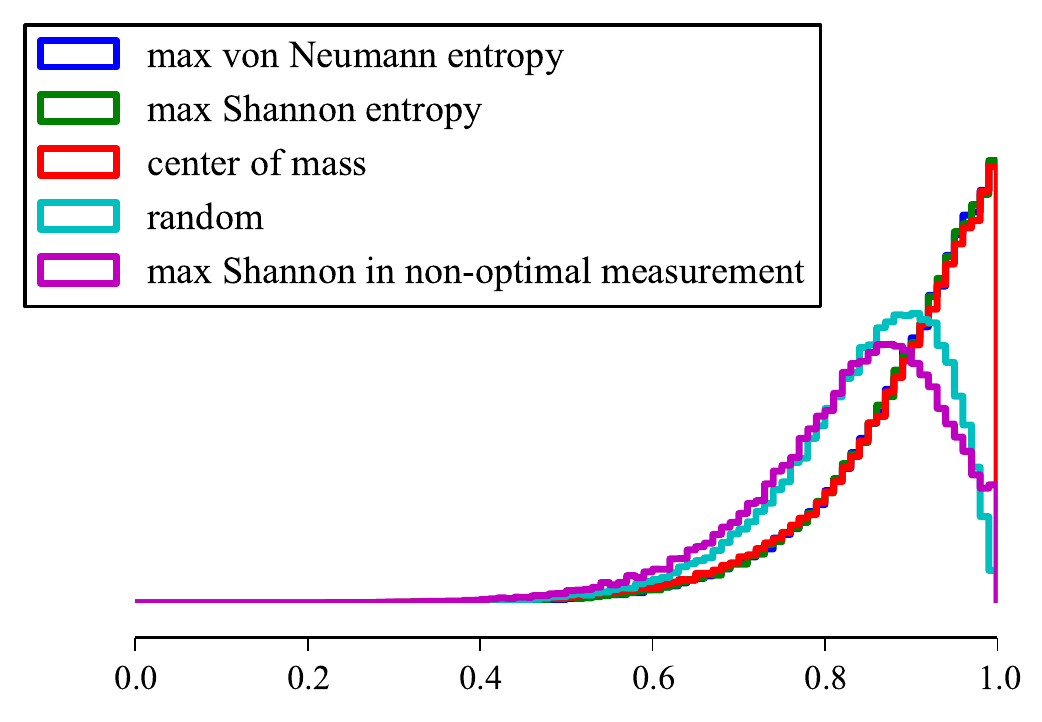}
  \caption{Sampling uniform on eigenvalue simplex, distance measured with fidelity.}
  \label{fig-estimatorcompare4Fid}
\end{figure}

\begin{figure}
  \centering
  \includegraphics[width=0.7\textwidth]{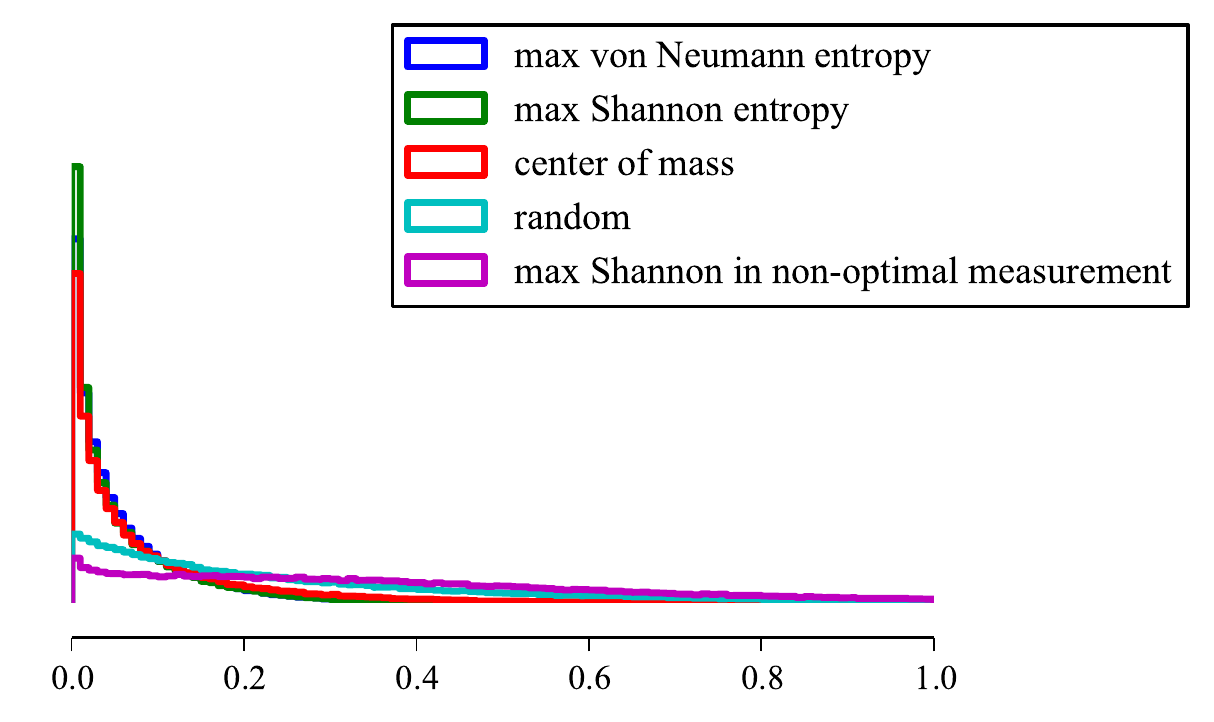}
  \caption{Sampling uniform on eigenvalue simplex, distance measured with quantum relative entropy.}
  \label{fig-estimatorcompare4RelEnt}
\end{figure}

\begin{figure}
  \centering
  \includegraphics[width=0.7\textwidth]{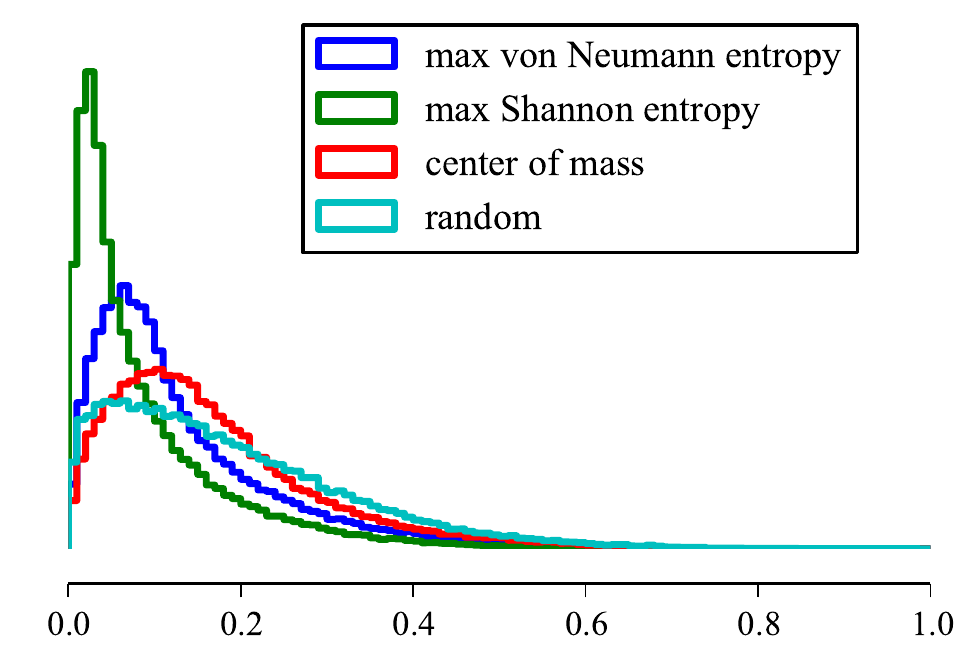}
  \caption{State of type $x\ketbra{\psi}{\psi} + \frac{1-x}{3}\mathbb I$ where $\ket{\psi}$ is random pure state, and $x=\sqrt\xi$ for uniformly drawn $\xi$ from $[0,1]$. Distance measured with Hilbert-Schmidt distance. Distribution of true states is uniform on purity, but does not sample fully from the space of states.}
  \label{fig-estimatorcompare7HS}
\end{figure}

\begin{figure}
  \centering
  \includegraphics[width=0.7\textwidth]{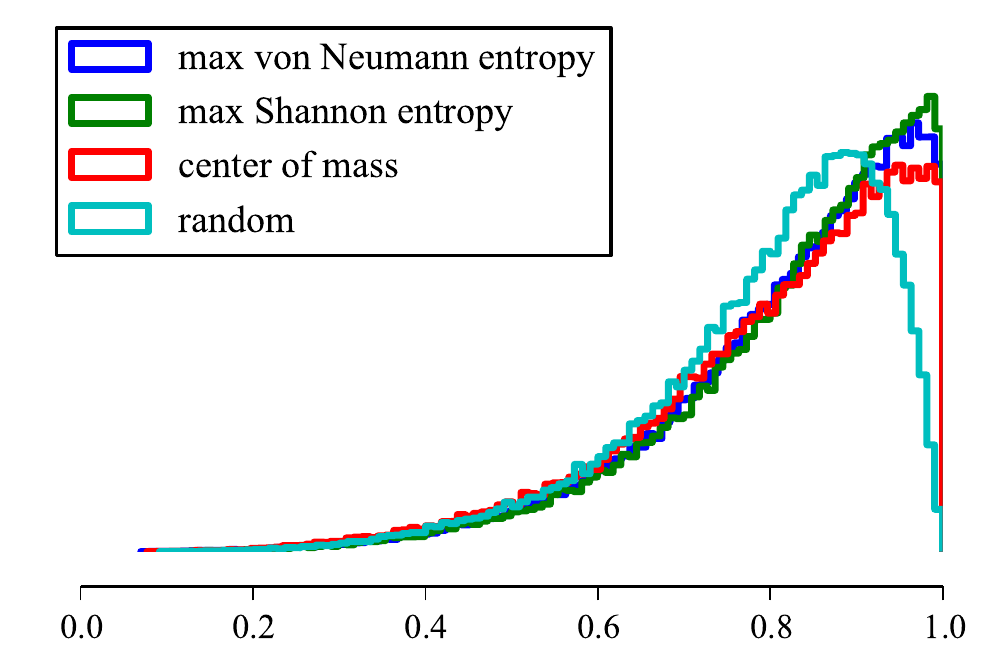}
  \caption{State of type $x\ketbra{\psi}{\psi} + \frac{1-x}{3}\mathbb I$ where $\ket{\psi}$ is random pure state, and $x=\sqrt\xi$ for uniformly drawn $\xi$ from $[0,1]$. Distance measured with fidelity.}
  \label{fig-estimatorcompare7Fid}
\end{figure}

\begin{figure}
  \centering
  \includegraphics[width=0.7\textwidth]{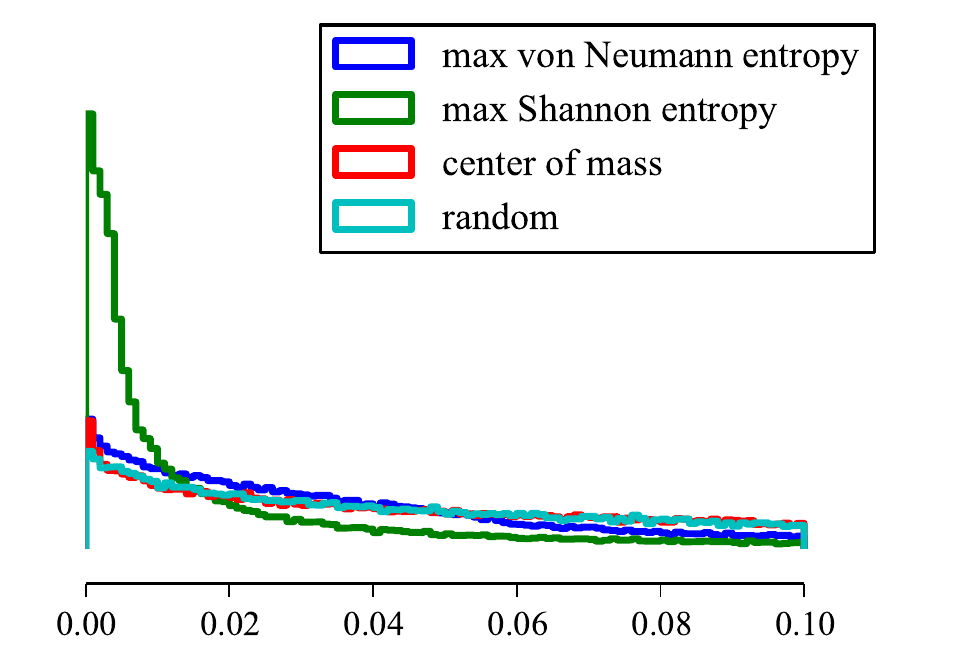}
  \caption{State of type $x\ketbra{\psi}{\psi} + \frac{1-x}{3}\mathbb I$ where $\ket{\psi}$ is random pure state, and $x=\sqrt\xi$ for uniformly drawn $\xi$ from $[0,1]$. Distance measured with quantum relative entropy.}
  \label{fig-estimatorcompare7RelEnt}
\end{figure}

The results figures show that MvNE, MSE in the unmeasured MUB, COM estimators to be competitive. Compared to these three, the random estimator, and MSE in another arbitrary future measurement do not perform well at all. Additionally, the maximum entropy methods performs somewhat better than the center of mass estimator. 

When the two bases of the MUB set are unmeasured, the results show similar pattern. The general conclusion we can make is that if we do not want to care about a future measurement, maximizing von Neumann entropy or calculating the center of the mass of the permissible region is a good way to obtain a competitive estimator. If one wants to select a measurement however, then it is better to choose a measurement with better performance. 

\begin{figure}
  \centering
  \includegraphics[width=0.7\textwidth]{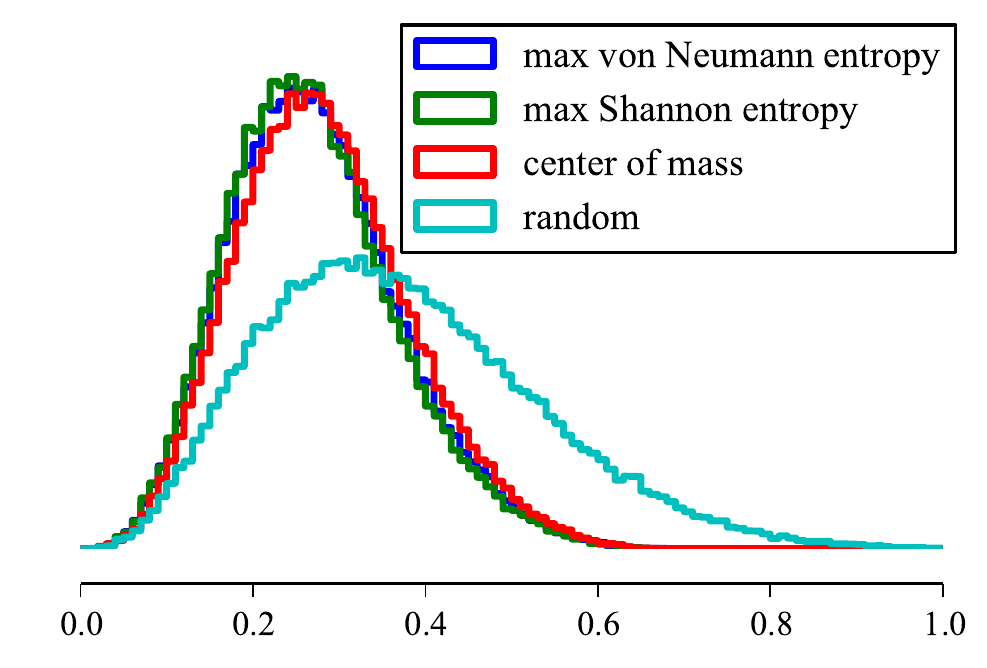}
  \caption{100000 true state sampled according to uniformity on Hilbert-Schmidt distance, distance between true states and estimators measured by Hilbert-Schmidt distance. Case of two unmeasured bases.}
\end{figure}

\begin{figure}
  \centering
  \includegraphics[width=0.7\textwidth]{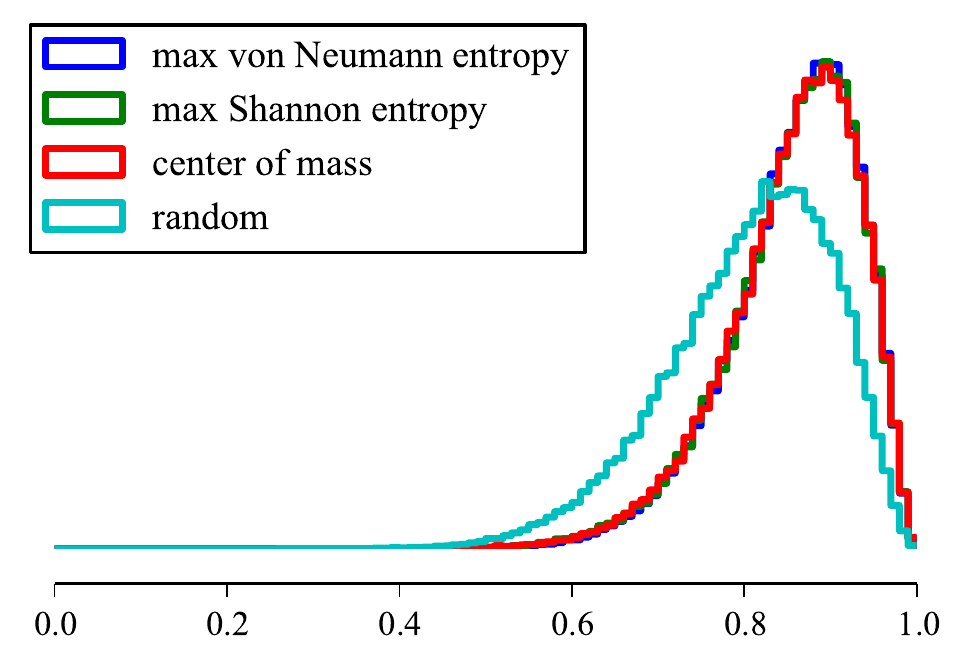}
  \caption{100000 true state sampled according to uniformity on Hilbert-Schmidt distance, distance between true states and estimators measured by fidelity Case of two unmeasured bases.}
\end{figure}

\begin{figure}
  \centering
  \includegraphics[width=0.7\textwidth]{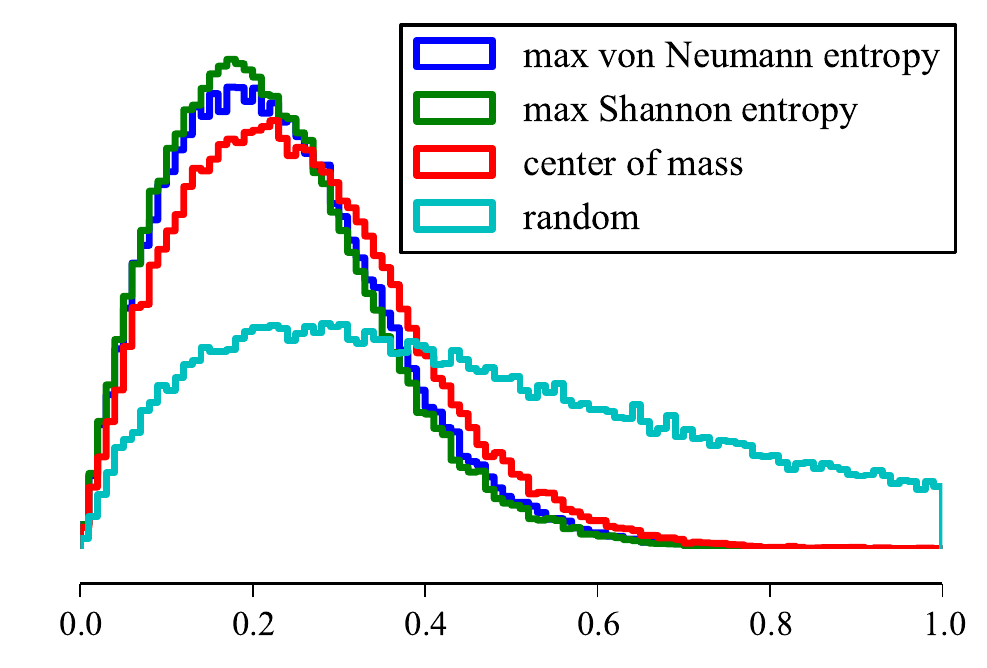}
  \caption{100000 true state sampled according to uniformity on Hilbert-Schmidt distance, distance between true states and estimators measured by quantum relative entropy. Case of two unmeasured bases.}
\end{figure}

\begin{figure}
  \centering
  \includegraphics[width=0.7\textwidth]{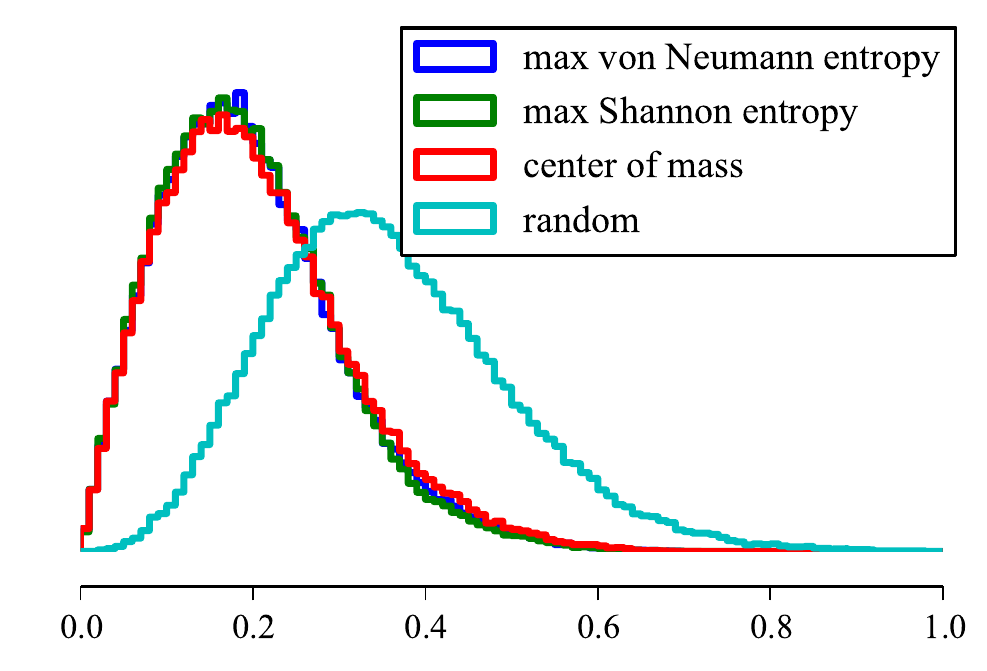}
  \caption{100000 true state sampled according to uniformity on eigenvalue simplex, distance between true states and estimators measured by Hilbert-Schmidt distance. Case of two unmeasured bases.}
\end{figure}

\begin{figure}
  \centering
  \includegraphics[width=0.7\textwidth]{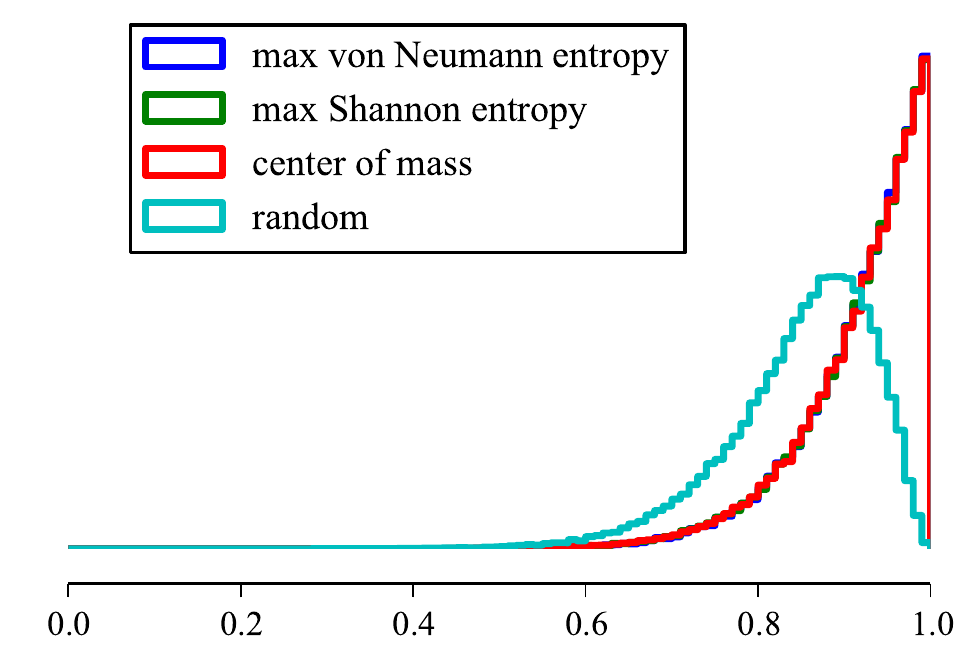}
  \caption{100000 true state sampled according to uniformity on eigenvalue simplex, distance between true states and estimators measured by fidelity. Case of two unmeasured bases.}
\end{figure}

\begin{figure}
  \centering
  \includegraphics[width=0.7\textwidth]{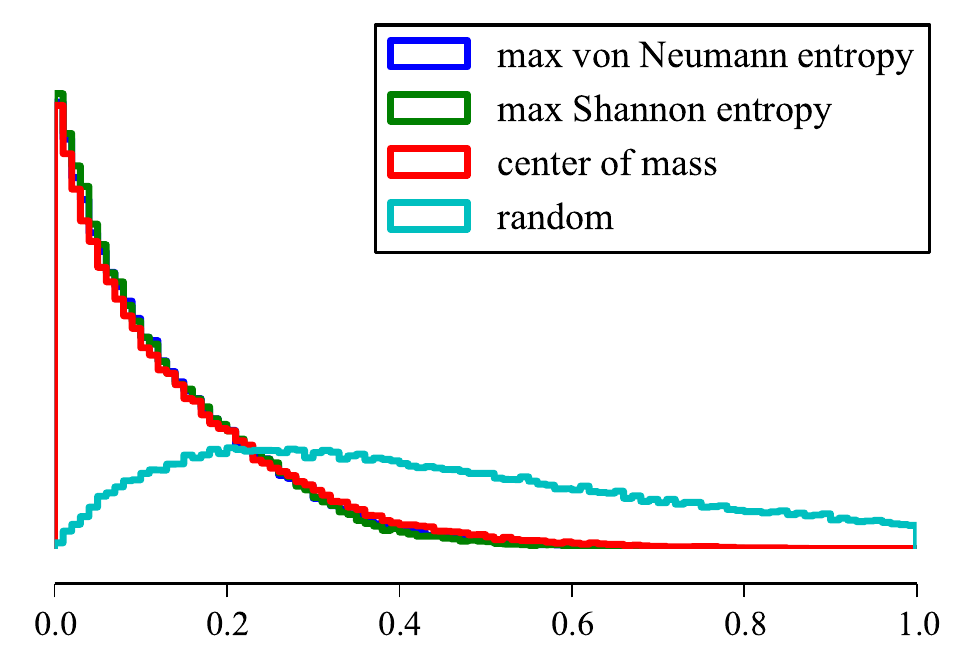}
  \caption{100000 true state sampled according to uniformity on eigenvalue simplex, distance between true states and estimators measured by quantum relative entropy. Case of two unmeasured bases.}
\end{figure}

\begin{table}
\centering
\begin{tabular}{@{}lccc@{}}
\toprule
\textit{HS sampling} & HS distance & Fidelity & Q. Rel. Entropy \\ \midrule
MvNE                 & 0.160       & 0.811    & 0.113           \\
optimal MSE          & 0.155       & 0.812    & 0.103           \\
COM                  & 0.179       & 0.802    & 0.161           \\
random               & 0.210       & 0.780    & 0.259           \\
non-optimal MSE      & 0.247       & 0.757    & 0.413           \\ \bottomrule
\end{tabular}
\caption{The average distance of estimators to sampled true states. One basis unmeasured}
\label{table-HSsample}
\end{table}

\begin{table}
\centering
\begin{tabular}{@{}lccc@{}}
\toprule
\textit{unif Eig sampling} & HS distance & Fidelity & Q. Rel. Entropy \\ \midrule
MvNE                       & 0.124       & 0.887    & 0.0664          \\
optimal MSE                & 0.118       & 0.887    & 0.0604          \\
COM                        & 0.134       & 0.882    & 0.0924          \\
random                     & 0.238       & 0.840    & 0.294           \\
non-optimal MSE            & 0.272       & 0.825    & 0.413           \\ \bottomrule
\end{tabular}
\caption{The average distance of estimators to sampled true states, one basis unmeasured.}
\label{table-unifEigSample}
\end{table}

\begin{table}
  \centering
  \begin{tabular}{@{}lccc@{}}
\toprule
\textit{HS sampling} & HS distance & Fidelity & Q. Rel. Entropy \\ \midrule
MvNE                 & 0.273       & 0.857    & 0.231           \\
optimal MSE          & 0.268       & 0.856    & 0.224           \\
COM                  & 0.283       & 0.855    & 0.256           \\
random               & 0.372       & 0.805    & 0.294          \\ \bottomrule
\end{tabular}
  \label{table-twobasisHS}
  \caption{The average distance of estimators to sampled true states. Two bases unmeasured.}
\end{table}

\begin{table}
\centering
\begin{tabular}{@{}lccc@{}}
\toprule
\textit{unif Eig sampling} & HS distance & Fidelity & Q. Rel. Entropy \\ \midrule
MvNE                       & 0.198       & 0.918    & 0.131           \\
optimal MSE                & 0.196       & 0.917    & 0.127           \\
COM                        & 0.203       & 0.917    & 0.143           \\
random                     & 0.356       & 0.849    & 0.562          \\ \bottomrule
\end{tabular}
\caption{The average distance of estimators to sampled true states. Two bases unmeasured.}
\label{table-twobasisUnifEig}
\end{table}

In particular, we see that a good Shannon estimator comes with a `best future measurement'. We have noticed a few things about this `best future measurement': It is the same as the unmeasured MUB. Also, the area of the permissible region is largest with respect to the probability simplex in this `best measurement'. The estimator combined with a prescribed best future measurement could guide experimenters to set up a specific experiment to complete the tomography of a system. Therefore, we have need to quantify what we mean by a `best future measurement'. What makes the `best future measurement', better than others?

\begin{savequote}[200pt]
	The Caterpillar was the first to speak. ``What size do you want to be?" it asked. ``Oh, I'm not particular as to size," Alice hastily replied; ``only one doesn't like changing so often, you know." ``I don't know," said the Caterpillar.
	\qauthor{Alice's Adventures in Wonderland, Lewis Carroll}
\end{savequote}
\chapter{Size of permissible region}
It was observed so far from our numerical results that the permissible region has size that depends on prior information and that it also varies with the choice of future measurement used --- being related to the Jacobian factor. It was also observed that the if two observers were to choose different future measurements to complete the tomography of the same system, the permissible region appears in different sizes relative to the probability simplex, even through the simplex is the same size for both of them. It suggests then that the size of the permissible region was an indication of the quality of the choice of the future measurements; we have already seen that while almost all future measurements allow for reconstruction of state, some are more useful when the estimator is derived for those measurement bases. 

In this chapter we discuss the problem of quantifying the area of the permissible region, so that we can compare these regions, and their significance to the question of optimal measurements.

\section{Euclidean $\mathbb R^n$ area}
Since we have justified using the uniform prior as a way of expressing our equal belief in each state within the permissible region as the possible true state, this prior defines a way to measure the area of the permissible region. In this case the measure would just be Euclidean on $\mathbb R^n$. It also implies that distances between states should be Euclidean. The advantage of using the Euclidean measure is that it is relatively easy to compute. However, the Euclidean measure does not take into account some of the characteristics of the transformation. In particular, the affine transformation described in \eqref{eq-probtransform} involves transformations of probability distributions, with $V$ being a stochastic matrix. As such the Euclidean measure may not appropiate, as the parameters that we are dealing with are probabilities, and the matrix elements for the affine transformations are transition probabilities.

Perhaps a better way to think about the measure of the $\Gamma$ is to consider the physical process by which $\Gamma$ is to be measured.

\section{Area based on experiments}
Since the coordinate parameters are probabilities, then it is inescapable to think about how we can identify the probability distribution for a random variable through an actual experiment. Since we can only measure probabilities through observations of a large number of independent, identical copies of the same random variable, any identification of the probabilities for the system must be accompanied by a suitable set of neighbouring probability distributions for which we are uncertain as to which of them might be the real one. Such uncertainty is always present in any real measurement, and is estimated as the `resolution' of the measuring device or `resolving power'. It also implies that the experimenter does not have to deal with infinitesimals, or continuous sets. Since his resolution for the set is always finite, it breaks up the measurable set into finite, countable pieces. In physics therefore, one does not really have a problem of measure, as the finite resolution of measurement means that we never encounter uncountably infinite sets. For example, a ruler is only capable of measuring lengths to a precision of 0.5 mm, but with a pair of vernier calipers we can improve this precision to 0.05 mm. The length of a side of a table is therefore based on how many `marks' we can count when we compare the side of the table against the ruler. With a pair of vernier calipers, we would be able to count more `marks', so the length of the table depends on the resolution of the measuring device in this sense. Something similar for probability distributions applies --- we can also think of observation counts as a kind of `resolution' to the measurement of probabilities. For example, if we have only tossed the coin about 10 times, we would not be able to tell if the coin has 0.48 probability of coming up heads, or 0.5 probability of doing so. But if we tossed the coin 10000 times, we might be able to do so. Therefore, the number of observation counts limit our ability to distinguish possible states of the coin. For a person tossing the coin only 10 times, he might only be able to distinguish 3 states of the coin: Fair, only heads or only tails. In this way, we can break up the continuous space of the probability simplex into a number of countable pieces. The idea is to count how many distinguishable states there are in $\Gamma$. This is an idea that was first used by Wootters \cite{wootters1981statistical} who used statistical distinguishability to define the distance between states.

How do we calculate the characteristic `resolution' for $N$ observations of outcomes for a measurement? In \cite{wootters1981statistical,braunstein1994}, this is calculated from the variance of the Gaussian distribution when the number of observations $N$ is large. While we will essentially obtain the same result, we can show the same result through the use of Bayesian inference methods, which are applicable even when $N$ is not sufficiently large.

Suppose we obtain some observations of measurement outcomes, $\{N_i\}$ which of course have $\sum_i N_i = N$. We now ask what is the probability that some state $\rho\in\Gamma$ is the true state, given the data $\{N_i\}$ obtained. To calculate this, we apply Bayes rule, and find that 
\begin{equation}
	\mathrm{prob}(\rho = \rho_\text{true}|\{N_i\}) \propto \mathrm{prob}(\{N_i\}|\rho)\mathrm{prob}(\rho=\rho_\text{true})\,,
\end{equation}
which says that our desired probability is proportional to the probability of the data being generated by $\rho$, as well as the probability that $\rho$ is the true state. In statistical theory, $\rho$ would be our hypothesis and $\{N_i\}$ is the evidence. We ask for the probability that the hypothesis is correct given the evidence, and Bayes rule tell us that the desired probability is given by the probability of evidence being generated by the hypothesis, and the probability the hypothesis being true amongst other competing hypotheses. The former is easily computed, and is known as the likelihood, while the latter has to be assigned by hand and as we have mentioned before, constitutes the prior. Suppose that the likelihood of the evidence being produced under the hypothesis is small, then the probability that the hypothesis being true given that we have observed this evidence will be small too. Suppose that the prior probability of the hypothesis being true amongst many competing hypotheses is low, then the probability of hypothesis being true will be small too, and furthermore, this is independent of the evidence. This has the amusing consequence that if we believe a friend to be absolutely trustworthy, then no finite amount of evidence to the contrary will make us believe otherwise. Of course Bayes rule is no longer necessary if he tooses the coin an infinite number of times.

Therefore, a state is a sufficient explanation for the data $\{N_i\}$ if its posterior probability is sufficiently large. This is crucial, because if we are to blindly consider the likelihood only, we will put into consideration false hypotheses. In this case I am refering to parameters which cause the reconstructed density matrix to have negative eigenvalues. It is because of the prior $\eta(\Gamma)$ that we assign to the probability simplex, that causes us to \emph{never} consider a density matrix with negative eigenvalues as the explanation for the data we obtain, even if the data seem to imply otherwise, due to statistical fluctuations. \cite{simpleminimax}

A state with maximum posterior could be an estimator for the data we obtained, but there is an associated neighbouring set $\gamma$ of the states that all have similar posterior to the maximum posterior state. This is in the same vein as the smallest credible region in \cite{shang2013optimal}, in which a region of states around the estimator is defined as the level of uncertainty due to finite $N$. That is, if we compare two competing hypotheses, and find both to be almost equally good in explaining the evidence, we should not favor one over the other. For quantum states, if both have comparable posterior as the true state, then they must be indistinguishable to the experimenter. His/Her data must be unable to say clearly as to which might have been the true state.
That is to say, if the log posterior odds,
\begin{equation}
	\label{eq-posteriorodds}
  \left|\log \frac{\mathrm{prob}(\rho_1|\{N_i\})}{\mathrm{prob}(\rho_2|\{N_i\})}\right| < 1\,,
\end{equation}
we say the we cannot distinguish $\rho_1$ from $\rho_2$, which is equivalent to saying that their posterior probability are within a factor of 10 of each other. This number is somewhat arbitrary, but then again measurement units has always been rather arbitrary; for example, we need to decide the meter in terms of a platinum bar, or define the start and end point of the centigrade scale. The factor to take into consideration is whether this definition allows for good measurement of the things that we want to measure. 

Now, under the assumption of the flat prior on $\Gamma$, $\eta(\Gamma)$ for the simplex, we assume the prior probability of $\rho_1$ being the true state is the same as the prior probability for $\rho_2$ to be the true state. Then we can see that the ratio of posteriors become only a ratio of likelihoods. If for some reason we believe $\rho_1$ to be 4 times as likely as $\rho_2$ to be the true state, then we should put this into equation \eqref{eq-posteriorodds} appropriately. Until then, we will assume the ratio to be 1.

For multinomial distributions of the observed outcomes, the ratio of likelihoods for two probability distributions $p$ and $q$ with some data $\{N_i\} = N\{f_i\}$ is
\begin{equation}
	\left|\sum_i Nf_i\log\frac{p_i}{q_i}\right| \,.
\end{equation}

Since we are considering the neighborhood $\gamma$ of the maximum posterior state (now maximum likelihood), we can set one of the states $p = f$. The details of this step can be filled in by considering the Lagrange multiplier method applied to the maximization of the likelihood function under the constraint of constant $N$. Technically, we are done here. By computing $\gamma$ as the set of the states with likelihood that is of some ratio with respect to the MLE, the details of which could be found in \cite{shang2013optimal}, we obtain a finite region that can be used a measure for $\Gamma$. By counting the number of the $\gamma$ that could be used to cover $\Gamma$ without overlap, we obtain the area of the $\Gamma$ that depends on `units' $N$.

A more tractable form for the measure could instead be found by considering a very small neighbourhood about the MLE. Within such a small neighborhood, we can apply Taylor expansion about the peak of the log likelihood function,
\begin{equation}
	\log \mathcal L \approx N\sum_i f_i\log f_i -\sum_j \frac{N}{2f_j}(q_j - f_j)^2\,.
\end{equation}
Therefore, near the MLE, the likeihood function of the data is approximately Gaussian. Here we note that the same Gaussian form was used as the starting step for analysis in \cite{wootters1981statistical,braunstein1994}, while we recognize this as an approximation. There, if the Central Limit theorem fails to hold, their analysis cannot continue, while we can still proceed with our basic notion of using $\gamma$ as a measure even when $N$ is small. However, the essential spirit for the underlying idea is the same: we use the uncertainty present within our measurement process to define a measure for the set $\Gamma$. The covariance matrix of the Gaussian we have derived is diagonal, so we can think of the Gaussian as an ellipse with characteristic lengths $\sqrt{\frac{2f_j}{N}}$. The number of distinguishable states is then simply a matter of counting:
\begin{align}
	\label{eq-countingdistance}
	\sqrt{\frac{N^3}{2^3}}\int\limits_{\substack{\Gamma\\p_1+p_2+p_3=1}} \frac{\D p_1\,\D p_2\,\D p_3}{\sqrt{p_1 p_2 p_3}}\,.
\end{align}
In the above few steps, we have made connections to several existing ideas. One is that we see a familiar connection to relative entropy between probability distributions as a measure of distinguishability. By expansion of the likelihood function around the peak, we obtain a metric for counting sizes of probability spaces. Therefore, two, we see the familiar expression of Jeffrey's noninformative prior again, but now as way of counting the number of distinguishable states. But it should be cautioned that this does not imply that we use this measure to define our prior, rather its the other way around: we used the flat prior to define this measure.

Finally, by a change of change of variables $\xi_i = \sqrt{Np_i}$, we see that the condition of unit probability becomes the condition of radius $\sqrt{N}$ on the positive octant of a sphere. Therefore, changing variables again to angular variables $\theta,\phi$, we find that the area of the permissible region $\Gamma$ is
\begin{equation}
	\int_\Gamma \sin\theta\,\D\theta\,\D\phi
\end{equation}
while the distance between two states is simply just the angular separation $\D\psi$. Then for $\vec\zeta=(p_1,p_2,p_3)$, $\vec\chi=(q_1,q_2,q_3)$, the angular separation is simply $\arccos(\sqrt{\vec\zeta}\cdot\sqrt{\vec\chi})$. But then we recognize \cite{wootters1981statistical} that the expression in the argument of the arccos function as the fidelity between probability distributions! 

While these measures and ideas have been in quantum information for a while now, they have been used in another context, in a communication channel for sending quantum states. In that context, we are given a promise of a finite set of quantum states that would be sent, and the receiver knows this set. The receiver just does not know which state amongst the set would be sent. In our context of tomography however, what we are proposing now is to view the situation as having a \emph{continuous} set of states to distinguish. By performing a experiment and making measurements, we are trying to distinguish the true state from the rest of the set of states. Why is this necessary? Theoretically, different measurement bases do not hinder our ability to reconstruct the state. However, physically, the fact that uncertainty is always finite, means that we cannot distinguish probability distributions perfectly. Therefore there exists a best measurement basis that we can use, such that the number of distinguishable states in the permissible region is as large as possible. Compare this with the notion of fidelity between mixed states (appendix). The fidelity between mixed states is defined as the minimum fidelity over all possible measurements that be made to distinguish the two states. In this sense the experimenter picks the best measurementment to use.

Therefore, one might try to define \emph{the} unique distance between states as the maximum distance possible using the best measurement. For our case, \emph{the} unique area of $\Gamma$ might be defined as the maximum area possible using the best measurement. This requires the maximization of area or distance under different measurements, and we deem this to be a possible future numerical work. As for the unique distance between quantum states, the authors of \cite{braunstein1994} managed to derive the Bures distance, performing the maximization through applying an inequality to obtain upper bound. While this is a nice theoretical result, it does not guarantee that the upper bound is the maximum, only that the maximum is at least given by the upper bound. Some numerical work is required here, to answer some possible questions --- Could the best measurement be just the complete MUB set or can come from a general measurement (POM)? If the best measurement could be found, then we also have a natural parameterization of $\rho$ such that computation of areas and distances would simply be in terms of the probabilities of this parameterization. This could be a relatively easier way to compute the Bures distance (known to be difficult to compute, because fidelity between mixed states is difficult to compute), if a connection to Bures distance can be confirmed.

Numerically, Monte Carlo method was used to evaluate the integral of equation \eqref{eq-countingdistance}, randomly sampling points on the simplex based on the Dirichlet distribution with alpha parameters $\alpha_i=0.5$. The ratio of accepted points to total sampled points multiplied by a factor is the answer. The factor required is simply the surface area of the positive octant of the unit sphere $\pi/2$. If we have to deal with more unmeasured bases, we can again use perform Monte Carlo again, but now multiplying by a factor of $(\pi/2)^m$ for $m$ unmeasured bases. If we are not working with qutrits or if we are working with more general measurements with more outcomes, then the factor needs to be accordingly adjusted --- the surface area of the hyperoctant is required then.

\section{A relook at the estimator performance}
Having justified the use of this particular way of measuring areas and distances, we can recompute the estimator performances. We will sample true states of 3 types: highly mixed states, rank 2 states, and pure states. We will also look at the ratio of distance to the square root of the area of the permissible region, which causes any dependence on $N$ to cancel out.

\begin{figure}
	\centering
	\includegraphics[width=0.7\textwidth]{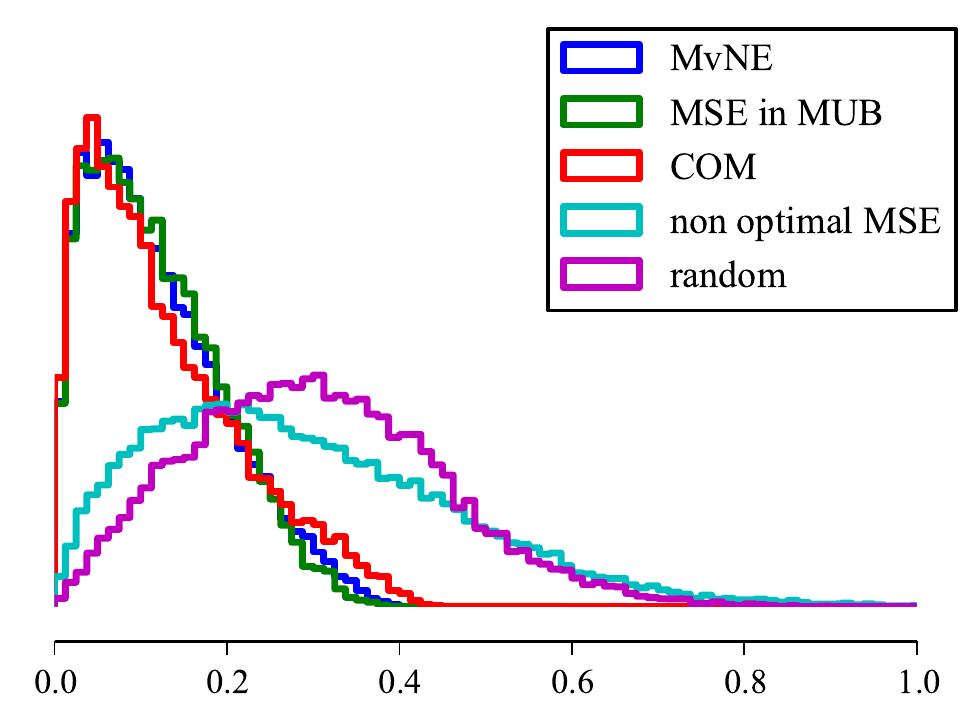}
	\caption{Distribution for ratio of distance to square root of area, for highly mixed states. This means a uniform sampling of states with purity less than 0.5. We simulate one basis unmeasured.}
	\label{fig-highlymixed}
\end{figure}

\begin{figure}
	\centering
	\includegraphics[width=0.7\textwidth]{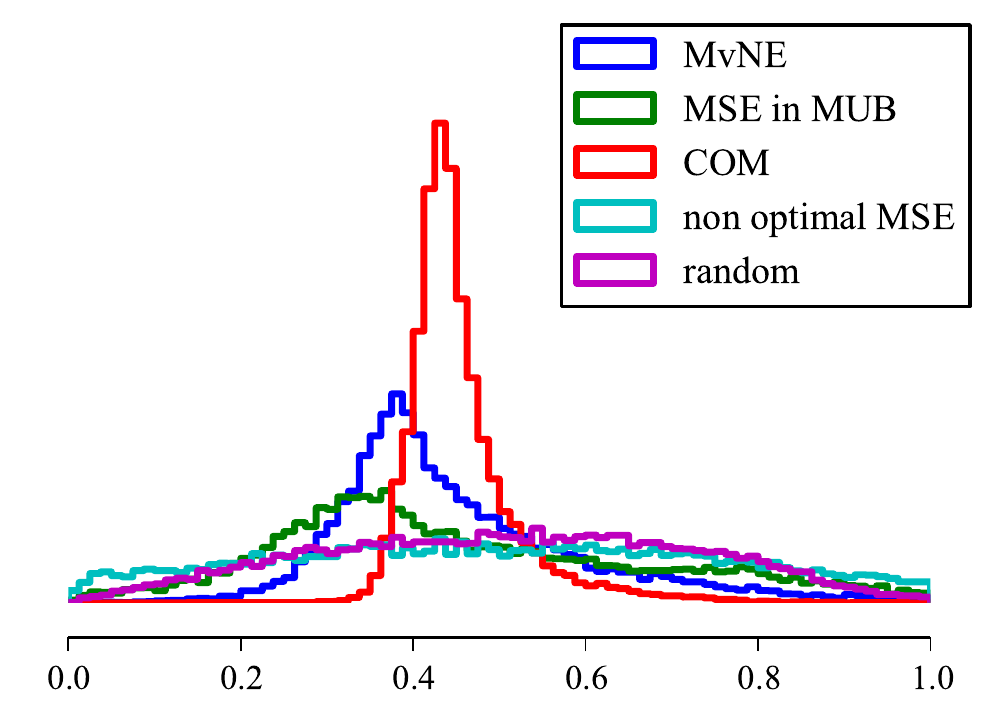}
	\caption{Distribution for ratio of distance to square root of area, for rank 2 states. This means a uniform sampling of states with eigenvalues $(0.5,0.5,0)$ to $(1,0,0)$. We simulate one basis unmeasured.}
	\label{fig-rank2}
\end{figure}

\begin{figure}
	\centering
	\includegraphics[width=0.7\textwidth]{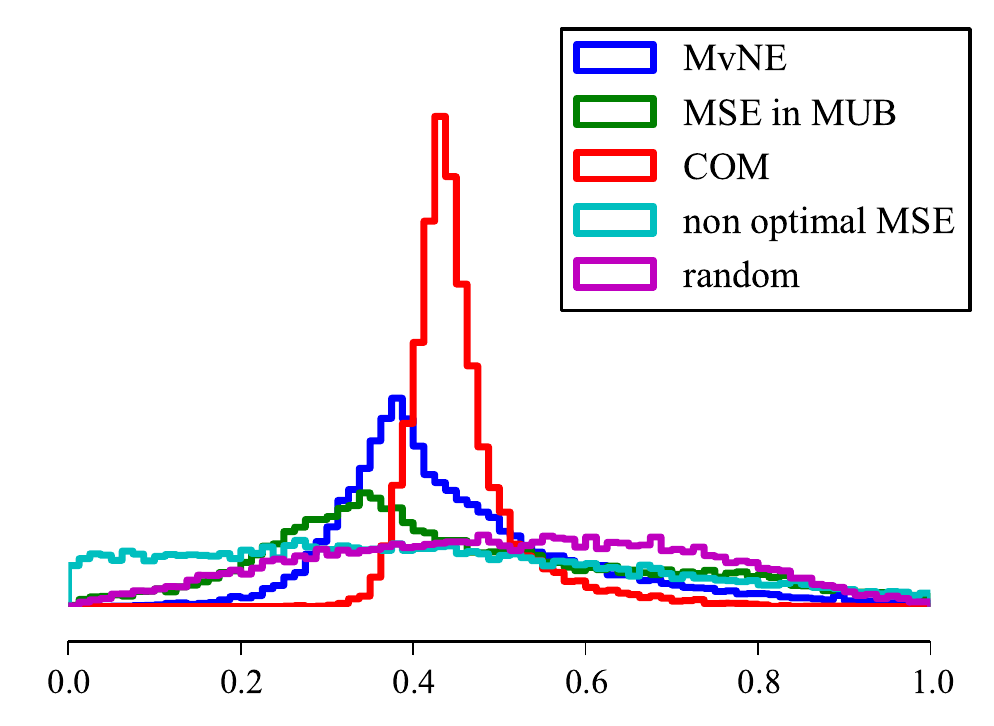}
	\caption{Distribution for ratio of distance to square of area, for pure states. Random samples are based on the Haar measure. We simulate one basis unmeasured.}
	\label{fig-distancepure}
\end{figure}

The conclusion remain similar, that maximum entropy methods and center of mass estimators remain competitive. However, we can also observe something extra. With highly mixed states (figure \ref{fig-highlymixed}), the competitive estimators perform overwhelmingly; which is not quite so surprising since the competitive estimators produce predictions near the completely mixed state. This is compounded by the fact that only highly mixed states gives rise to large permissible regions. In such cases where the quantum state is essentially classical-like, we expect classical inference techniques to work well --- predicting that the unknown random system gives uniform probabilities for outcomes is a reliable answer. Any other kind of strategy like random picking or maximizing the `wrong' entropy produces non-optimal predictions.

With rank-deficient states such as rank 2 states or pure states, the results are quite different (figures \ref{fig-distancepure}, \ref{fig-rank2}). The most noticeable feature is the relatively sharp peak for the COM estimator. It is centered roughly about 0.42. This is rather unsurprising if we remember that the true state in this case lies on the boundary of the permissible region. Therefore the COM, which tends to be at the center of the region, would almost always be around half of the diameter of the permissible region to the true state --- it is a consistent estimator in this case. What could be more interesting is that the maximum entropy methods show a greater variation in their distances to the true state; but on average, they perform somewhat better than the COM. The ordering of their average distance is MSE in MUB $>$ MvNE $>$ COM. As hypothesized previously, it could be that when the permissible region is small, the MSE frequently lie on the boundary of the permissible region. In some sense, if the MSE is selected correctly for the right future measurement, it could be much closer to the true state than other estimators. Compare this to the non-optimal MSE. The non-optimal MSE shares a fate similar to the random estimator. Their distribution of distances is essentially uniform compared to the competitive estimators.

We have also noted previously that when the true state is rank deficient, the permissible region has a tendency to be small. Sometimes, when the permissible region is too small, the algorithms can fail to locate an estimator. In the course of producing figures \ref{fig-highlymixed}, \ref{fig-rank2}, \ref{fig-distancepure}, 20000 true states were sampled for each type of the states. Amongst these samples, the algorithm failed 24\% of time for rank deficient states, compared with the 4\% failure rate for highly mixed states. It was found by further numerical trials, that the reason for algorithm failure was when $\Gamma$ shrinks to become a point. When this happens, optimization algorithms can come as close to this point as we like, but due the numerical error, they never get the point exactly. As such they end up with an estimator that has negative eigenvalues. The Monte Carlo methods simply fail to find $\Gamma$ altogether. It must be remarked that in these cases, while predicting the estimator should be very easy --- the point $\Gamma$ itself, locating this point exactly is by no means easy for the experimenter. Like the numerical algorithm, the experimenter will have unavoidable uncertainty, for his region of uncertainty is now larger than $\Gamma$, and so he will scratch his head because all his estimators based on his gathered data will not be physical states. But in this case he should be delighted instead, for this is an indication that the system is in a state of high purity.

\section{Possible applications}
With this notion of computing the area of the permissible region, we can compare the areas for the permissible region under different parameterizations. We have observed in numerical trials that the permissible region is largest for measurements from the set of mutually unbiased bases. Using this notion of computing the area, maximizing the area of the permissible region is akin to the situation where the experimenter searches for the basis to measure such that he can distinguish the most number of states.

Therefore, it is a possible that for quantum systems with dimensions in which the full set of mutually unbiased bases is unknown, the experimenter can search for optimal measurements sequentially through optimizing the permissible region at each step. However, the usefulness or veracity of this is unknown to us. More analysis and numerical explorations is required to verify its utility and truth. However, it could be similar to other numerical searches for the MUB, using the distances between bases as the criterion \cite{raynal2011mutually}.

The measurement of area offers a different perspective on the problem of MUB, as we have here an experimental consideration of why it makes sense to select the MUB to measure. We can think of a basis that is mutually unbised to previously measured bases as the best measurement to make in next step, until we complete the tomography of the state.

\chapter{Closing Remarks}
We have looked at several ways to compute estimators for the case of unmeasured bases for a qutrit, discussing their rationale, and numerically computing the state as the estimator. For optimization, a gradient ascent algorithm was used, with the critical piece of the algorithm as the maximization of the minimum eigenvalue in order to guide iterates into the permissible region. With Bayesian mean estimators, we have seen how a prior probability distribution is required as a way to give weight the states in the permissible region. We have given some justification for using the uniform prior distribution on the simplex of probabilities parameterizing the state, and argued that without further evidence that this is unsatisfactory, we can continue to use on the basis of simplicity. The numerical computation of the mean proceeds by a simple Monte Carlo method.

The different methods of producing estimators are compared by simulating true states and incomplete measurements in order to apply the different estimator strategies. The distance of the true state to the estimator is then compared, and we have found, by different distance measures, that the estimators we have discussed are almost equally competitive. On the other hand, estimators produced from the use of another future measurements are not quite as competitive, similar to the random estimator.

From the point of view of tomography, we have another perspective as to why the MUB set is useful --- it is possibly the set that allows the experimenter to better distinguish the true state from the set of possible states. For a 3 dimensional quantum system, where the MUB set is known, we can confirm that this is indeed the best set of measurement to perform tomography. However, in other dimensions where the MUB is not analytically known, we can still fall upon the physical intuition of distinguishability in order to find an optimal or almost optimal set of bases.

The results of this project also gives us more understanding about incomplete tomography, such that we can now proceed to consider more complicated situations: for example, if instead of probabilities of measurement bases or numbers for the density matrix, we are given expectation values for certain observables such as the mean energy of the quantum system, how do we now take these kinds of constraints into account?

\appendix
\chapter{Miscellania}
In this chapter are collected the odds and ends that do not fit into the main narrative for this project, but are things that are picked up by the author as useful concepts to know, or small observations that were interesting (but not a major discovery), or otherwise, simply numerical details that are too technical for all except another person who wishes to pick up the project. Here is the scaffold used during the construction of this project.

\section{Mutually Unbiased Bases, MUB}
Two bases are said to be mutually unbiased \cite{bengtssonMUB1,durt2010mutually} if the transition probability of any state in one basis to any state of another basis is uniform for all such pairs. 

For an orthonormal set of basis states $\{b_i\}$ in the Hilbert space, it was observed \cite{durt2010mutually} that the corresponding basis vectors $\Lambda'_i$ forms a regular simplex that spans an $d-1$ plane. Since the dimension of the space is $d^2-1$, we can then fit at most $d+1$ totally orthogonal planes into it. So the number of elements of the basis set $\{\Lambda_j\}$ to parameterize $\rho$ is $d(d+1)$, comprising of $d$ subsets of $d+1$ kets: each forming a orthogonal measurement basis for the Hilbert space. 

Observe that $\Lambda_i\cdot\Lambda_j = 0$ implies that
\begin{equation}
\label{eq-mubcondition}
\left|\bracket{e_i}{e_j}\right|^2 = \frac{1}{d}\,.
\end{equation}
This suggests that if we can find kets that satisfy \eqref{eq-mubcondition}, we can find vectors in the vector space that are orthogonal. In particular, we can arrange for the aforementioned planes to be orthogonal. Then, if any two kets from different measurement basis subsets that satisfy \eqref{eq-mubcondition}, the two measurement basis are said to be \emph{mutually unbiased}. Given a state corresponding to a ket from a basis, the outcome probabilities for the states of a basis unbiased to it is uniform.

In terms of more familiar concepts in quantum mechanics, the set of mutually unbiased bases constitute a full set of complementary observables, much like how momentum and position of a particle is complementary. The question of their existence and construction of a maximal set of mutually unbiased bases (MUB) for any dimension $d$ is generally difficult and would take us far afield from quantum physics. Instead, we will simply state a set of MUB for the qutrit $(d=3)$ that will be referred to in latter sections.

Given a computational basis set enumerated as $\{\ket{i}\}$, we find that the Fourier basis is automatically unbiased with respect to the computational basis. The Fourier basis is given by
\begin{equation}
\ket{f_j} = \sum_k \omega^{jk}\ket{k} \text{ with } \omega = \mathrm{e}^{-2\pi \mathrm{i}/3}\,.
\end{equation}
Its components can be represented by the columns of a matrix
\begin{equation}
F = \frac{1}{\sqrt 3} \begin{bmatrix} 1 & 1 & 1 \\ 1 & \omega & \omega^2 \\ 1 & \omega^2 & \omega \end{bmatrix}\,.
\end{equation}

To find more kets that are unbiased with respect to the columns of $F$, we search \cite{bengtssonMUB2} for vectors of the form $v = (1,\Exp{\I\alpha},\Exp{\I\beta})/\sqrt{3}$ such that
\begin{equation}
\left| F\cdot v \right| = \frac{1}{\sqrt 3} \begin{pmatrix} 1 \\ 1 \\ 1\end{pmatrix}\,,
\end{equation}
which leads to a set of 3 equations to solve, leading to $\alpha,\beta = 0,\frac{\pi}{3},\frac{2\pi}{3}$. The maximal set of MUB components is given here as columns of 4 matrices:
\begin{equation}
 \label{eq-mubmatrix}
\begin{bmatrix} 1 & 0 & 0\\ 0 & 1 & 0\\ 0 & 0 & 1 \end{bmatrix}\,,\frac{1}{\sqrt 3} \begin{bmatrix} 1 & 1 & 1 \\ 1 & \omega & \omega^2 \\ 1 & \omega^2 & \omega \end{bmatrix}\,, \frac{1}{\sqrt 3} \begin{bmatrix} 1&1&1\\ \omega&\omega^2 & 1\\ \omega & 1 & \omega^2 \end{bmatrix}\,,  \frac{1}{\sqrt 3} \begin{bmatrix} 1 & 1 &1 \\ \omega^2 & \omega & 1 \\ \omega^2 & 1 & \omega \end{bmatrix} 
\end{equation}

In terms of the MUB, the statistical operator has a convenient representation:
\begin{equation}
  \rho + \mathbb I = \sum_{i,j} p_{i,j}\ketbra{i^{(j)}}{i^{(j)}} 
\end{equation}
where $\ket{i^{(j)}}$ refers to the $i$th column of the $j$th matrix in \eqref{eq-mubmatrix} and $p_{i,j}$ refers to the corresponding probability that we will find outcome $\ket{i^{(j)}}$ given $\rho$.

\section{An observation regarding the permissible region in a special case}
In the special case where the off-diagonal matrix elements are precisely pre-determined but the diagonal elements are unknown, we have discussed the permissible region must lie within the region enclosed by 3 curves: the principle 2 by 2 subdeterminants. That is for
\begin{equation}
	\begin{bmatrix} x & a & b \\ a^* & y & c\\ b^* & c^* & z \end{bmatrix}
\end{equation}
the permissible region must lie within the intersection of
\begin{align}
	xy - |a|^2 \geq 0\qquad
	yz - |c|^2 \geq 0 \qquad
	xz - |b|^2 \geq 0\,.
\end{align}
Therefore, by symmetry concerns, it is more likely to find parameters with nonnegative eigenvalues for the corresponding density matrix along the lines that join the center of the simplex $(1/3,1/3,1/3)$ to the points $(1/2,1/2,0)$, $(1/2,0,1/2)$, $(0,1/2,1/2)$. Such an observation aids in locating the permissible region in the special case, especially in cases when the permissible region has small area relative to the simplex.

\section{Convex Optimization problems}
The standard \cite{boyd2004convex} convex optimization problem is stated as:
\begin{subequations}
\begin{align*}
\operatorname*{minimize}_{x\in\mathbb R^n}\, &f_0(x) \\
\text{subject to  } &f_i(x)\leq 0\,,\qquad i = 1,\,\cdots ,\, m.\\
\hphantom{subject to  } & Ax = b \WITH A\in \mathbb{R}^{p\times n}\,, \text{rank } A = p < n\,,
\end{align*}
\end{subequations}
where the form the problem is stated so as to make it clear that the $m$ inequality constraints are of the convex type, and equality constraints can be put into affine form. Naturally, $f_0(x)$ is a convex function. Naturally, maximization of a concave function is equivalent to minimizing its negative.

\subsection{Karush-Kahn-Tucker conditions}
A necessary condition that is satisfied by the sought after optimal point $x^\star$ is that the following equations are statisfied:
\begin{subequations}
\begin{align}
\nabla \mathcal{L} = \nabla f_0(x^\star) + \sum_i^m \lambda_i^\star \nabla f_i(x^\star) + A^T\mu^\star = 0\,, \\
Ax^\star = b\,,\qquad f_i(x^\star) \leq 0\,,\qquad 1,\,\cdots,\,m \,,\\
\lambda^\star \succeq 0\,,\\
\lambda_i^\star f_i(x^\star) = 0\,,\qquad i = 1,\,\cdots,\,m\,,
\end{align}
\end{subequations}
where we see that these conditions are simply a generalization of the method of Lagrangian multipliers.

\subsection{Variable step-size in gradient ascent}
In order to promote faster convergence, the step-size between iterates is calculated by a back-tracking line search algorithm. After the direction $\vec v$ is determined in the main optimization algorithm, the update step from current iterate $\vec x_0$ to next one $\vec x_1$ requires calculation of the step size $t$ in $\vec x_1 = \vec x_0 + t\vec v$. 

The step size is calculated as follows. Keeping in mind that we want to maximize a function $f$, we check the following repeatedly, starting with $t=1$, while
\begin{equation}
	f(\vec x_0 + t\vec x_1) < f(\vec x_0) + \alpha t|\vec v|^2\,,
\end{equation}
set $t = \beta t$. This condition is known as Armijo's rule. Here, $\alpha$, $\beta$ are tunable parameters, set through trial and error to achieve the best result. In our course of the project, it was found that $\alpha=0.25$ and $\beta=0.5$ were sufficient. The idea behind the Armijo's rule is to ensure sufficient increase in the objective function by taking steps that are not too large.

Another trick required for the optimization was that once the iterates were within the permissible region, the back-tracking line search could step outside the region again. This situation has to be caught by the algorithm manually, by setting the function values outside the permissible region to a large negative value.

\subsection{Rationale for choosing log barrier method}
Initially, the Lagrangian written for Shannon entropy was chosen as the objective function for gradient ascent optimization. However, it was found that convergence could not be achieved reliably as optimal points frequently lay on the boundary of the permissible region. This point was not maximum point but a saddle point in terms the parameters and the multipler. Moreover, due the numerical error, the solution found by the algorithm could be a non-feasible solution even though close to the theoretical optimum. 

The log barrier method is a method that prevents iterates from leaving the permissible region. Iterates can reach as close to the barrier as possible, but cannot step over, as the function value becomes infinity at the boundary itself. The only drawback is that the Hessian becomes ill-conditioned near the optimum at the boundary.

\section{Random states and unitary matrices}
In this section we deal with some background theory about random quantum states, and random measurement bases. For discrete random variables, the statement `picking a random outcome uniformly' is clear and unambiguous. However, for continuous random variables, the same statement is now dependent on how the domain is measured. Counting the size of the discrete objects is unique, but counting the size of a continuous space is not so! 

\subsection{Random points in parameters of a vector space}
We will start with what we are familiar with: We know how to sample uniformly from $\mathbb R^n$. Suppose we can parameterize physical states with $(u_1,u_2,\cdots,u_n)$. If there are no other constraints on the parameters, then a uniformly sampled state is simply just a uniform sample on $\mathbb R^n$. If there are linear constraints, we can eliminate them. For example we have $\sum_i u_i = 1$, then we can rewrite as
\begin{equation}
(u_1,u_2,\cdots,u_n) = \vec r + \sum_i^{n-1} v_i\vec v_i\,,
\end{equation}
where $\vec v_i$ are suitable vectors, and $v_i$ are the corresponding coefficients. $\vec r$ is a displacement vector. Now we can sample uniformly from $(v_1,\cdots,v_{n-1})$. But what if we have nonlinear inequality constraints? For example, consider a two dimensional case $(u_1, u_2)$, and we have the constraint that $u_1^2 + u_2^2 \leq 1$. We can consider the region as embedded in $\mathbb R^2$, and utilize rejection sampling. That is, we uniformly sample on $\mathbb R^2$, and reject points $u_1^2 + u_2^2 > 1$. 

On the other hand, if we have a nonlinear equality constraint, then things become a little harder. \cite{diaconis2013sampling} If we have a parametrization: a bijective function $f$ such that $f: [0,1]^n \rightarrow \Omega$ where $\Omega$ is a sub-manifold of dimension $k$, $k < n$, then we can calculate the derivative (Jacobian) matrix $J$, where its elements are 
\begin{equation*}
[\mathrm D f]_{ij} = \mathrm D_i f_j\,.
\end{equation*}
The Jacobian determinant is defined as $|\mathrm{det}\,[ (\mathrm Df)] |$. The volume $V$ of the sub-manifold is determined through the metric, which is related to the Jacobian, 
\begin{equation}
\idotsint |\mathrm{det}\,[ (\mathrm Df)^{\mathrm T} (\mathrm Df)] | \D u_1\cdots\D u_n\,.
\end{equation}
Therefore $V$ must be a uniform random variable. Suitably normalized, we should sample using $J(u_1,\cdots,u_n)$ as distribution. Numerically, this can be achieved in various ways, by inverse transform method or rejection sampling for smaller dimension cases, and Markov chain Monte Carlo for large dimension distributions. 

\subsection{Parametrizing unitary matrices}
A unitary matrix $U$ satisfying $U^\dagger U = \mathbb I$ has orthonormal columns which can represent a set of orthonormal quantum pure states. We can parametrize $U$ by a set of orthonormal quantum pure states. There are many ways of deriving this parametrization, but they are all equivalent in one way or another. 

Because the columns/rows of the unitary matrix are complex-valued vectors with length 1, the modulus of the individual matrix elements correspond to hyper-spherical coordinates. A convenient parametrization for $SU(3)$ is the following from \cite{recursivefactorizationunitary}:
\begin{align}
\label{eq:unitaryfactorization}
&\mathrm{diag}(\Exp{\I\phi_1},\Exp{\I\phi_2},\Exp{\I\phi_3}) \begin{bmatrix} 
\cos\theta_1 & -\sin\theta_1 & 0\\
\sin\theta_1\cos\theta_2 & \cos\theta_1\cos\theta_2 & -\sin\theta_2 \\
\sin\theta_1\sin\theta_2 & \cos\theta_1\sin\theta_2 & \cos\theta_2 \end{bmatrix}\times\nonumber\\
&\quad\mathrm{diag}(1,\Exp{\I\phi_4},\Exp{\I\phi_5}) \begin{bmatrix}
1 & 0 & 0\\
0 & \cos\theta_3 & -\sin\theta_3 \\
0 & \sin\theta_3 & \cos\theta_3 \end{bmatrix} \mathrm{diag}(1,1,\Exp{\I\phi_6})\,.
\end{align}
This is a recursive definition of the $U(n)$ in general, with the complex phases separated out from the real valued orthogonal part.  

According to what we have discussed in the previous section, a uniform sample of unitary matrices should come from the distribution defined by the Jacobian determinant for this parameterization. After some lengthy computation, we find that we should sample according to
\begin{subequations}
\begin{align}
\phi_i \sim \mathrm{Unif}(0,2\pi)\,, i = 1,2,\cdots,6\,,\\
\sin^4\theta_1 \sim \mathrm{Unif}(0,1) \,,\\
\sin^2\theta_2 \sim \mathrm{Unif}(0,1) \,,\\
\sin^2\theta_3 \sim \mathrm{Unif}(0,1)\,.
\end{align}
\end{subequations}

Properly, if one has a continuous group that one would like to measure the size for, one talks about the Haar measure on such a group. Observe that $U(n)$ forms a Lie group, so in order to talk about integrals such as $\int \D U$ we need a way to assign a invariant volume under the group operation. The Jacobian determinant of the above is a Haar measure for $U(3)$. 

Numerically, uniform random samples of $U(n)$ according to the Haar measure can be generated by a random $n$ by $n$ complex matrix with independent entries sampled from normal distribution $\mathcal N(0,1)$, for both real and imaginary parts. Gram-Schmidt orthogonalisation procedure is then applied to produce orthonormal columns. Numerically, one applies the QR factorization, factoring the input matrix into unitary $Q$ and triangular matrix $R$. However, modern numerical packages do not use Gram-Schmidt steps to perform the factorization, using instead Householder reflections which do not produce the correct final distribution for the eigenvalue of $U$ (Explanation deferred as it requires a lengthy exposition on groups, see \cite{mezzadri2006generate}). The remedy is simply to take $U = QD$ where $D = \mathrm{sgn}\,\mathrm{diag}(R)$. 

\subsection{Random quantum states}
According to the previous section, random pure states can be found from the columns of a Haar random unitary matrix. The question of a random quantum mixed state however, is more complicated. One way is to make use of the parametrization of unitary matrices: By writing $\rho = U^\dagger \rho_d U$, we can generate random matrices by generating $U$ and diagonal $\rho_d$ separately. However there are some caveats to watch out for --- the first is that the diagonal can be parametrized in $n-1$ variables due to unit trace condition of the density matrix. Sampling the values of the diagonal matrix is the same as sampling uniformly on a simplex. The second caveat is that while we used $n^2$ parameters to characterize a unitary matrix, we only need $n^2 - n$ variables for the unitary component in the density matrix. This is because according to the factorization we have before for unitary matrices, we can rewrite the density matrix (in the qutrit case) as
\begin{equation}
\rho = \cdots \mathrm{diag}(\Exp{-\I\phi_1}, \Exp{-\I\phi_2},\Exp{-\I\phi_3})\,\rho_d\,\mathrm{diag}(\Exp{\I\phi_1}, \Exp{\I\phi_2},\Exp{\I\phi_3})\cdots\,,
\end{equation}
where the dotted parts are the other factors on the unitary matrix as according to \eqref{eq:unitaryfactorization}. Observe that the relevant factors displayed here are diagonal matrices that commute, so in fact the diagonal matrices containing phases cancel out here, eliminating $n$ redundant variables (3 for the qutrit case). Therefore, the total number of real valued variables required to parametrize a complex hermitian matrix is $n^2 - 1$, as can be confirmed by straightforward counting. Accordingly, this parametrization gives us a metric  $J^{\mathrm T}J$ from the Jacobian matrix $J$, and volume element $|\mathrm{det}J^{\mathrm T} J|$. In particular, the metric tensor gives a way to measure distances for the manifold of mixed quantum states.

Alternatively, we can also express traceless Hermitian matrices as vector, as we have done so in the beginning chapter. As a side note, a slightly way is to parameterize the entire $\rho$ by a symmetric frame, known as a SIC-POM, or SIC-POVM in literature. However, because the mapping of states to vectors is not bijective, the Jacobian of this linear parameterization does not suggest a way to sample from the space of states, but rather from the space of $\mathbb R^{(d^2-1)}$ instead.

Based on these two ways of parameterizing the state, we can sample random states accordingly. \cite{inducedmeasure}

\subsubsection{Distribution on eigenvalue simplex}
\begin{figure}[h]
  	\centering
  	\includegraphics[width=0.7\textwidth]{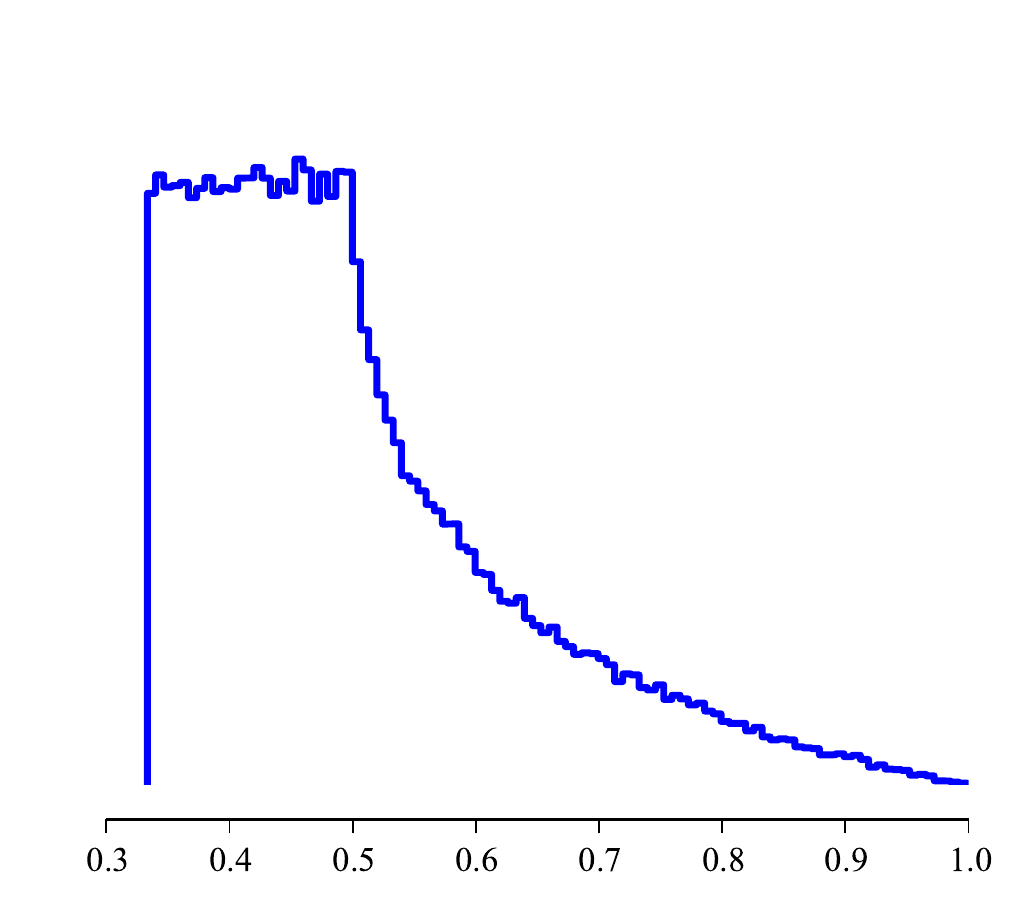}
	\caption{The distribution for purity of states sampled according to uniformity on the eigenvalue simplex. This shape is not unexpected given that the simplex is a triangular flat plane with discrete symmetry, while states of constant purity lie on a circle.}
	\label{fig-puritydist-unifeig}
\end{figure}

If we think of the state as a unique mixture of orthogonal states, then we can sample the unitary part from Haar measure, and the eigenvalue part from a measure on the eigenvalue simplex. A general distribution on this simplex is the Dirichlet distribution. This is a family of probability distribution controlled by a vector of parameters. Figure \ref{fig-puritydist-unifeig} shows the distribution for purity of states if we sample uniformly on the eigenvalue simplex.

\subsubsection{Distribution on probabilities in MUB parameterization}
\begin{figure}[hb]
  \centering
  \includegraphics[width=0.7\textwidth]{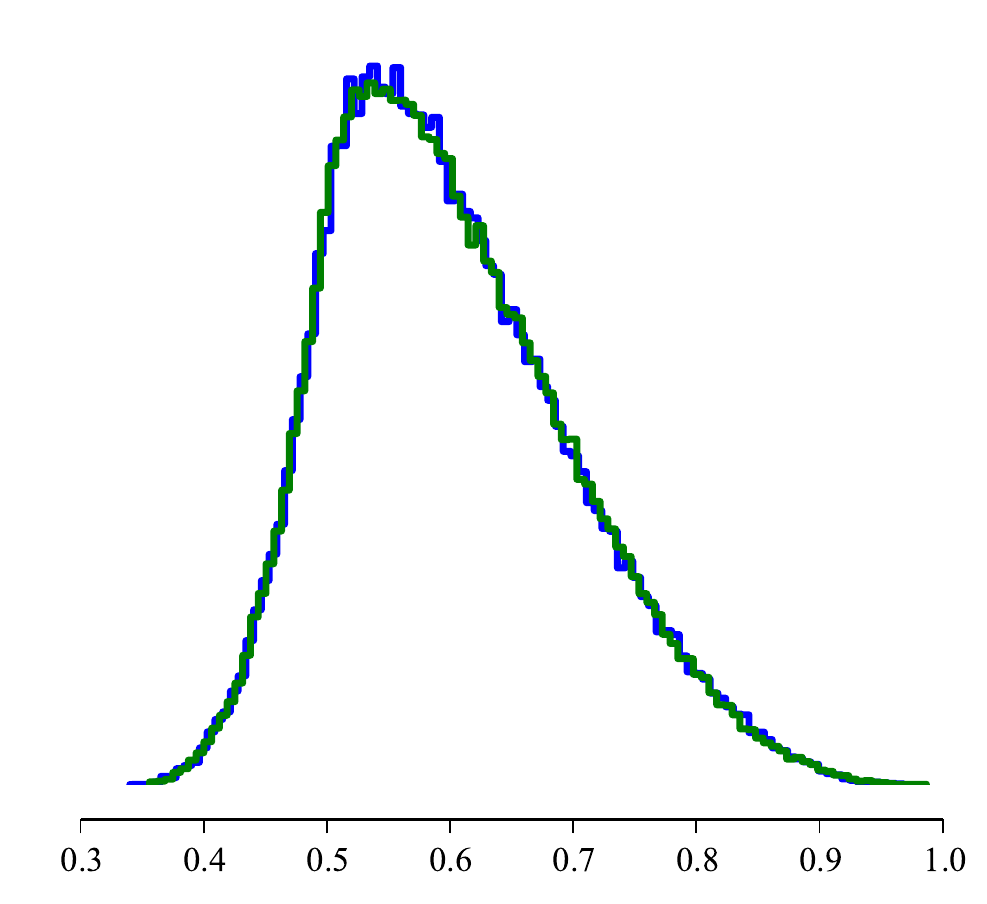}
  \caption{Purity distribution for states sampled according to method drawing matrix elements with normally distributed real and imaginary components. Superimposed are states sampled according to using accept/reject procedure on probabilities for the MUB parameterization of the qutrit, drawn uniformly on the probability simplex.}
	\label{fig-puritydist-HSsicpom}  
\end{figure}

\begin{figure}
  	\centering
  	\includegraphics[width=0.7\textwidth]{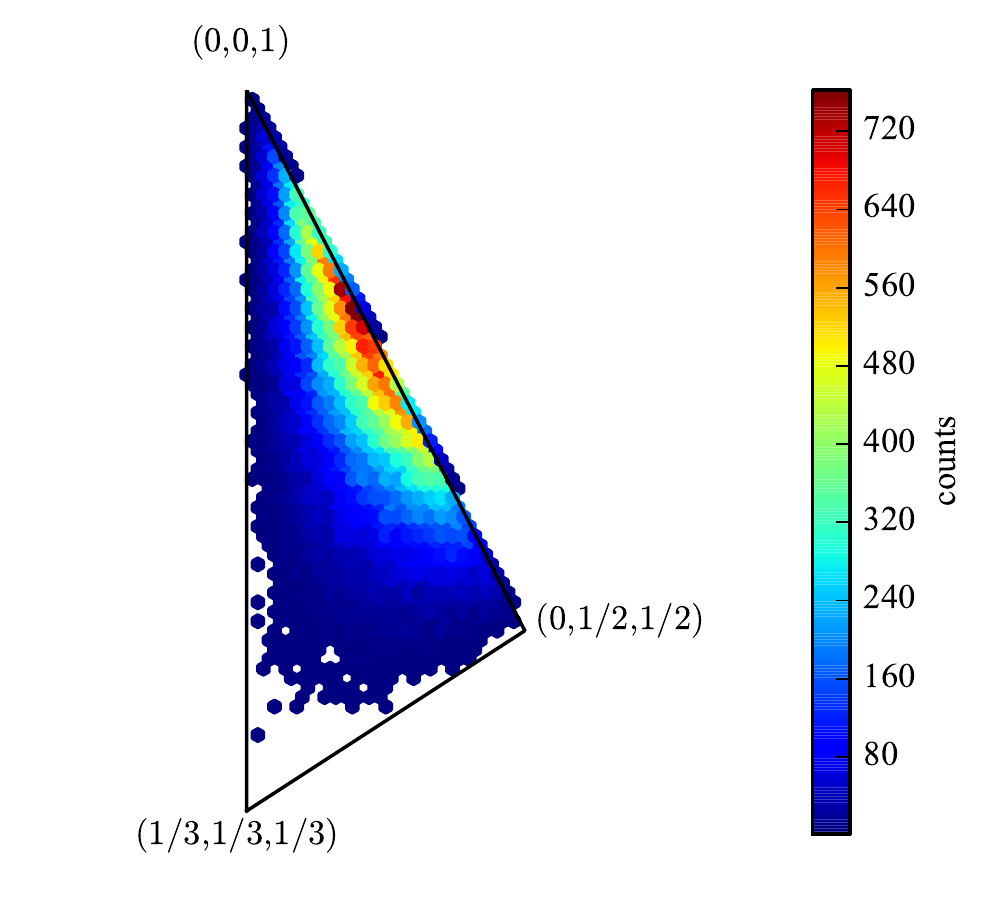}
	\caption{Distribution of eigenvalues for 100000 states sampled according to uniformity of HS distance.}
	\label{fig-eigdistri-HS}
\end{figure}

We can also think of randomly choosing points for probabilities. If we pick uniform points on the probability simplexes corresponding to the 4 bases in the MUB representation, we obtain another way of choosing random states. It was found by numerical trial and error, that this way of sampling is equivalent to choosing a random matrix $A$ with its real and imaginary elements drawn from standard normal distribution, and then taking $\rho = AA^\dagger/\Tr{AA^\dagger}$. This is a simpler method of sampling because when choosing probabilities for the MUB parameterization, we need to reject points that do not correspond to physical states. It was further shown in \cite{zyczkowski2011generating} that this simple method of sampling is equivalent to taking a random pure state of a larger dimension, and the partial tracing out these extra dimensions. The purity distribution for such a method of sampling is shown in figure \ref{fig-puritydist-HSsicpom}, and the distribution of eigenvalues on the simplex is shown in figure \ref{fig-eigdistri-HS}. Both show how the rank 2 states have greater tendency to be sampled, while pure states and fully mixed states are hardly sampled.

\subsection{Distance measures}
One of the canonical measures of distance between quantum states, is fidelity. It is a measure of similarity; for two pure states it is $|\ketbra{\psi}{\phi}|$. It remains intuitively easy to understand for the case of a pure state and a mixed state: $\sandwich{\psi}{\rho}{\psi}$ being the transition probabilities of the orthogonal states making up rho to the state $\ket{\psi}$. For the distance between two mixed states however is a lot less intuitive. It is given by
\begin{equation}
F(\rho,\rho') = \Tr{\sqrt{\sqrt{\rho_1}\rho_2\sqrt{\rho_1}}} = ||\sqrt{\rho}\sqrt{\rho'}||_1 \,.
\end{equation}
It was shown \cite{fuchs1994ensemble} that this expression is equivalent to the distance between probability distributions where the experiment chooses the best generalized measurement, POM, to distinguish the states.

While fidelity is not a proper metric, proper metrics can be derived from it, one of them being the Bures distance.
\begin{equation}
\mathrm d(\rho_1,\rho_2) = \sqrt{ 2(1 - F)}\,.
\end{equation}

Another canonical measure of distance is the Hilbert-Schmidt distance, which can be intuitively understood to be somewhat like Euclidean distance between matrices. The inner product in Hilbert space can be used to define a distance measure,
\begin{equation}
	d_\text{HS} = \sqrt{\Tr{(\rho-\sigma)^2}}\,,
\end{equation}
where we can see the resemblance to the Euclidean distance formula.

Finally, the relative entropy $\rho$ relative to $\sigma$, is given by
\begin{equation}
	\Tr{\rho(\log\rho - \log\sigma)}\,.
\end{equation}
This is similar to the definition of Kullback-Leibler divergence between probability distributions, and indeed if the states are commuting, the definitions becomes the same as the classical situation. Hence it is not surprising to find that the relative entropy is a measure of our ability to distinguish two quantum states, just like fidelity.

\section{Computation of boundary states of permissible region}
It may useful to know sometimes the rank deficient states that make up the boundary. For example, we may be interested in known whether a pure state exists on the boundary of the permissible region, after having made some measurements. Here we describe a simple algorithm to obtain a discretized approximation of the boundary.

We start with any state within the permissible region. So one way to obtain such a state is to take the MvNE, or run the maximization of minimum eigenvalue algorithm to find such a state.

Since the permission region is a closed, convex region, we can use angular parameters to parameterize. For example, for one unmeasured basis of the qutrit, the boundary is essentially just a single dimensional closed curve, which we can parameterize by $(\sin\theta,\cos\theta)$. For some $\theta_0$, we can calculate the boundary state for this parameter by essentially a method of bracketing. We already have state that is within the permissible region, given by some parameters $(u,v)$. Then we consider $(u,v) + \mu(\sin\theta,\cos\theta)$, by letting $\mu$ be sufficently large so that we have a non-physical state.

Once we have a line segment with one endpoint inside the permissible region, and the other endpoint outside the permissible region, then somewhere along the this line segment is the boundary. We simply carry out bisection search, using a Cholesky factorization as a fast criterion to check for positivity.

By random sampling of angles, we can obtain a discrete mesh of boundary states in this manner. A gradient ascent type of algorithm may be used to actively search a for particular state. For example, by minimizing von Neumann entropy on the boundary, we may find a pure state.
\backmatter
\bibliographystyle{apsrev-title}
\bibliography{fyp}

\end{document}